\documentclass[aps,prd,twocolumn,superscriptaddress]{revtex4-1}

\usepackage{longtable}
\usepackage{grffile} 
\usepackage{graphicx} 
\usepackage{amsmath}    
\usepackage{hyperref}   
\usepackage{subfigure}  
\usepackage{color}

\def\sh2{\hat s^2}
\def\th2{\hat t^2}
\def\uh2{\hat u^2}

\begin{document}


\title{Photon structure studied at an Electron Ion Collider}

\author{X. Chu}\email{xchu@mails.ccnu.edu.cn}
\affiliation{Key Laboratory of Quark and Lepton Physics (MOE) and \break Institute
of Particle Physics, Central China Normal University,\break Wuhan 430079, China}
\affiliation{Physics Department, Brookhaven National Laboratory, \break Upton, NY 11973, U.S.A.}
\author{E.C. Aschenauer}\email{elke@bnl.gov}
\affiliation{Physics Department, Brookhaven National Laboratory, \break Upton, NY 11973, U.S.A.}
\author{J.H. Lee}\email{jhlee@bnl.gov}
\affiliation{Physics Department, Brookhaven National Laboratory, \break Upton, NY 11973, U.S.A.}
 \author{L. Zheng}\email{liangzhphy@gmail.com}
 \affiliation{Key Laboratory of Quark and Lepton Physics (MOE) and \break Institute
of Particle Physics, Central China Normal University,\break Wuhan 430079, China}


\begin{abstract}
A future Electron Ion Collider (EIC) will be able to provide collisions of polarized electrons with protons and 
heavy ions over a wide range of center-of-mass energies (20 $\mathrm{GeV}$ to 140 $\mathrm{GeV}$) at an 
instantaneous luminosity of $10^{33} - 10^{34}$ cm$^{-2}$s$^{-1}$.
One of its promising physics programs is the study of the partonic structure of quasi-real photons. Measuring 
di-jets in quasi-real photoproduction events, one can effectively access the underlying parton dynamics of the 
photons. In this paper, we discuss the feasibility of tagging resolved photon processes and measuring the di-jet 
cross section as a function of jet transverse momentum in the range of $0.01<x_{\gamma}^{rec}<1$ at an EIC. 
It will be shown that both unpolarized and polarized parton distributions in the photon can be extracted, and 
that the flavor of the parton can be tagged at an EIC.
\end{abstract}


\maketitle

\section{Introduction}

The study of photons has a long history.
On the theoretical side, the first idea that energy can be emitted and absorbed only in discrete portions comes 
from Planck and was presented in
1901 in his successful theory describing the energy spectrum of black body radiation~\cite{Planck}.
Soon after Planck made his heuristic assumption of abstract elements of energy,
Einstein proposed that light can be considered as a flux of particles in 1905~\cite{EinsteinA}.
Many further experiments, beginning with the phenomenon of Compton scattering, validated Einstein's hypothesis 
that light itself is quantized.
In 1926 the optical physicist Frithiof Wolfers and the chemist G.~N.~Lewis referred to these particles as the 
notion of ``photon''.
Over the last century we have witnessed tremendous progress in our understanding of photons.
In quantum electrodynamics (QED), the photon mediates the electromagnetic force between charged objects.
As the gauge boson of QED, the photon is considered to be a massless and chargeless particle~\cite{VVKobychev} 
having no internal structure.
QED also incorporates the electron, which was the first elementary particle correctly identified as such.
The understanding of reactions involving these two particles spawned the theory of gauge interactions,
now thought to describe all observed (electroweak, strong and gravitational) interactions.
In spite of this long, distinguished history, there is one large class of photonic interactions about which only 
relatively little is known.
In any quantum field theory, the existence of interactions implies that the quanta themselves can develop a structure.
According to quantum chromodynamics (QCD), the photon is a superposition of a bare photon state which interacts only 
with electric charges, and a hadronic photon state.
If experimentally probed at very short distances, the intrinsic structure of the photon is recognized as a flux of 
quark and gluon components, quasi-free according to asymptotic freedom in QCD and described by the photon 
structure function~\cite{WalshTF,WSlominski,WittenE}.
For example, the photon can fluctuate for a short period of time into a charged fermion-antifermion pair,
$f\bar{f}$, carrying the same quantum numbers as the photon.
The lifetime of this fluctuation increases with the energy of the parent photon $E_{\gamma}$ and decreases
with the square of the invariant mass of the pair $M^{2}_{pair}$: $\Delta t \approx 2E_{\gamma}/M^{2}_{pair}$.
As a result, the photons will interact with hadrons (or other real photons) proceeding into two quite different ways.
The photon, as a whole particle, can couple directly to a quark in the struck hadron (direct process).
Alternatively, the photon can undergo a transition into a (virtual) hadronic state before encountering the target 
hadron (resolved process).
In this case a quark or gluon ``in'' the photon can react, via strong interactions, with partons in the struck 
hadron.
Then we can refer to the photon structure, which is a consequence of quantum fluctuations of the field theory. 

\begin{figure} 
\includegraphics[width=0.23\textwidth]{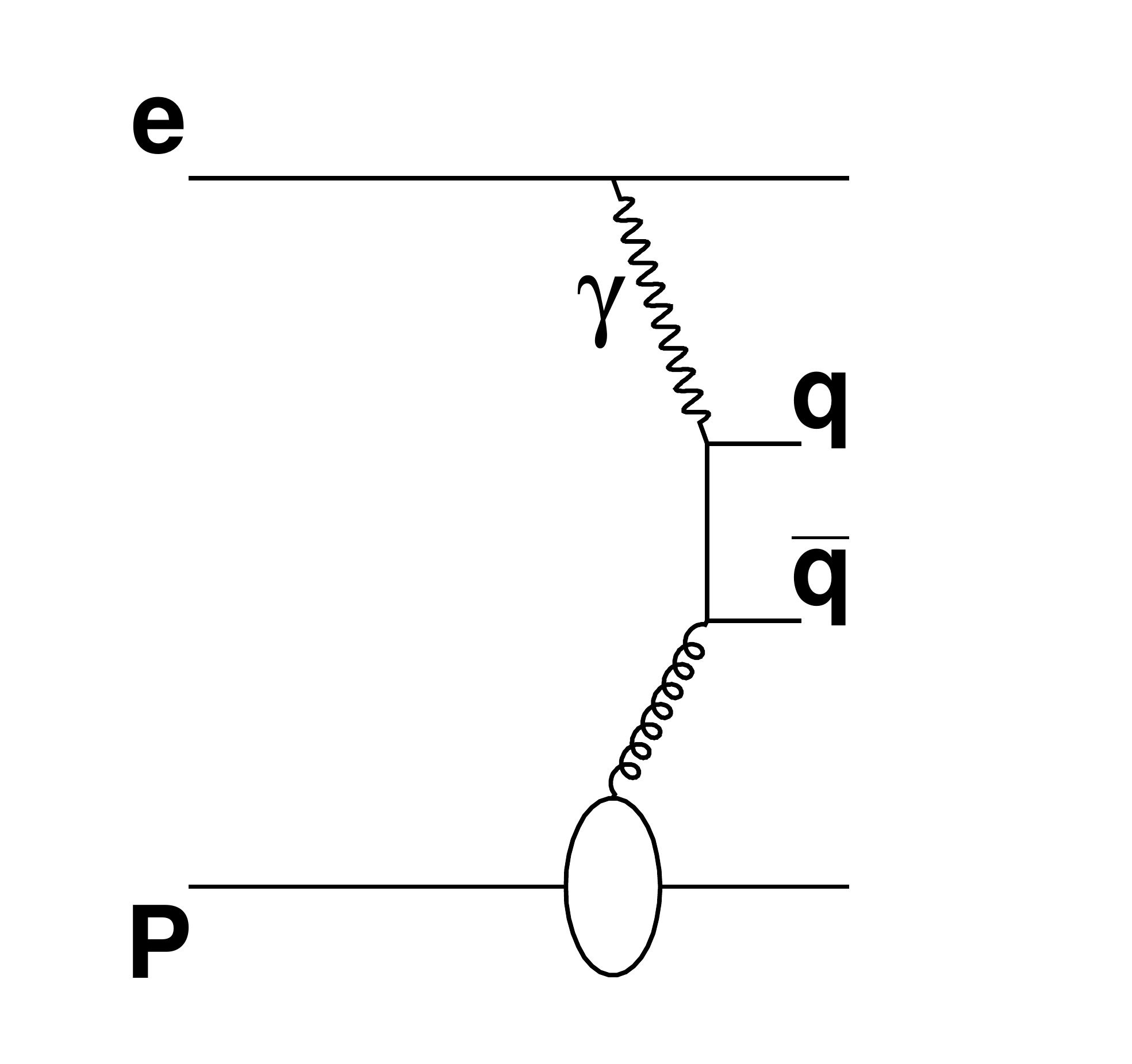}
\includegraphics[width=0.23\textwidth]{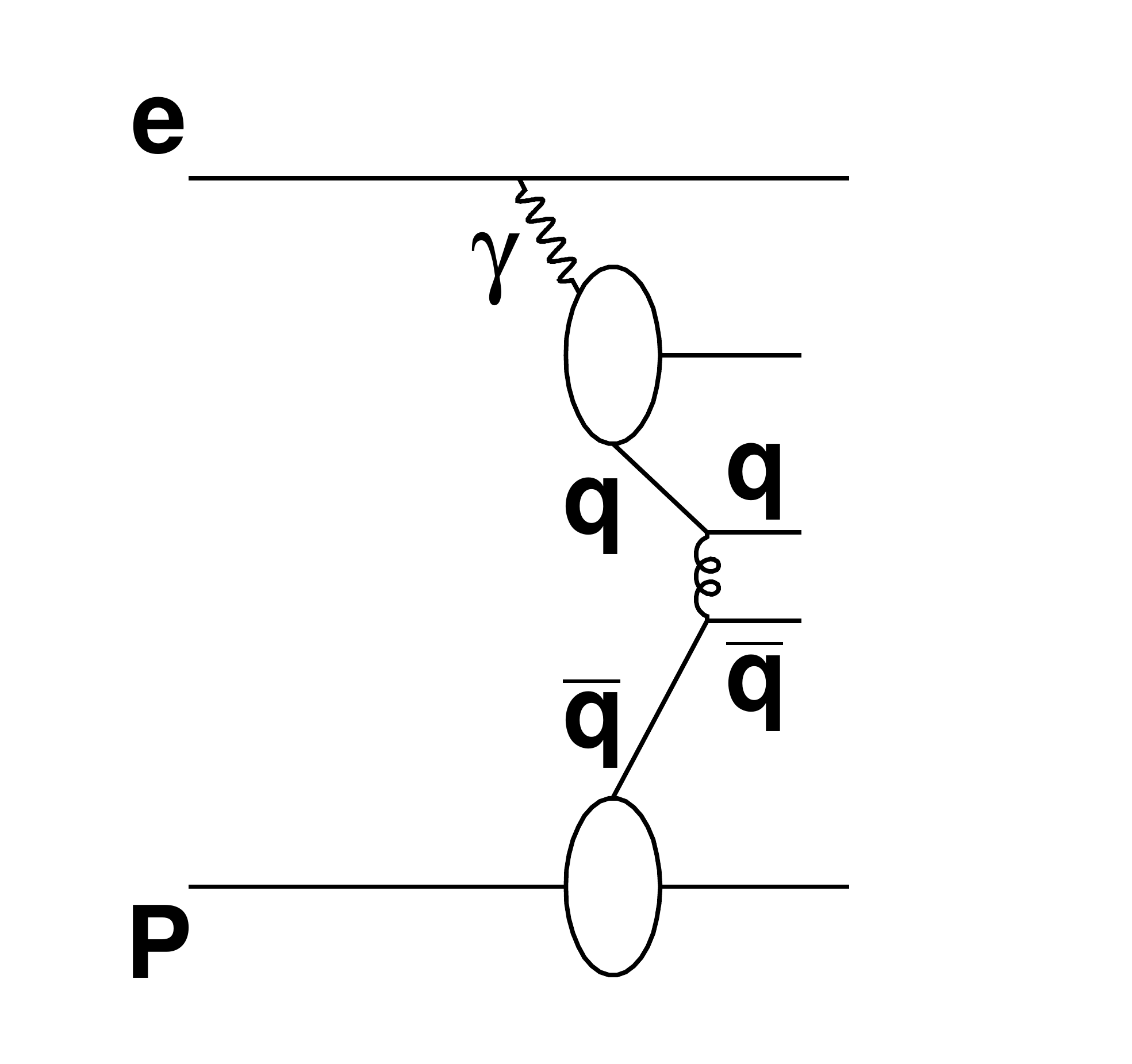}
\caption{Examples of diagrams for direct (left) and resolved (right) processes in electron-proton scattering.}
\label{fig: Feynman diagram}
\end{figure}

In the past decades, experimental progress to constrain the photon QCD structure has come mainly
from e$^+$e$^-$-collider experiments~\cite{F2,ArminBöhrer,ADeRoeck} and to some extent from the HERA
experiments~\cite{JMButterworth,PJBussey}.
The classical way to investigate the structure of the photon at e$^{+}$e$^{-}$ colliders is to measure the 
following process:
\begin{equation}\label{eq1.1}
e^{+}e^{-}\rightarrow e^{+}e^{-}\gamma^{\star}\gamma^{\star}\rightarrow e^{+}e^{-}X,
\end{equation}
proceeding via the interaction of two photons, which can be either quasi-real 
($\gamma$) or virtual($\gamma^{\star}$),
where X represents a pair of leptons or a hadronic final state.
In the collider HERA at DESY, 820 or 920 GeV protons collided with 27.5 GeV electrons or positrons,
with two general purpose detectors, H1 and ZEUS,  positioned at opposite interaction regions.
The high flux of almost on-shell photons which accompanied the lepton beam also provided a unique opportunity
to study the nature of the photon and its interactions.
Unlike in $e\gamma$ scattering, the photon structure is probed by the partons from the proton in the so-called 
photoproduction events in $ep$ collisions.
By tagging high transverse energy ($E_{t}$) jets~\cite{H1Collaboration1},
high-$p_{T}$ charged particles~\cite{H1Collaboration5} or heavy quarks~\cite{SFrixione} in photoproduction reactions,
Parton Distribution Functions (PDFs) of the photon can be constrained.
The interaction of electrons and protons at low virtuality is dominated by quasi-real photoproduction processes where 
the electrons scatter at small angles.
Such reactions proceed via two classes of processes, the so-called ``resolved" and ``direct" processes.
Examples of Feynman diagrams of these two processes are shown in FIG.~\ref{fig: Feynman diagram}.

In this paper, we perform a detailed study of the feasibility of measuring the photon structure via di-jets
at a future high-luminosity, high-energy Electron Ion Collider (EIC)~\cite{AAccardi}.
We demonstrate that, at a future EIC such as eRHIC at BNL, it is feasible to do a high precision extraction of 
photon PDFs
with an integrated luminosity of $\mathcal{L}=1 \, fb^{-1}$.
More importantly, an EIC also allows study of the polarized photon PDFs, as both the electron and proton beam 
can be polarized.
Table~\ref{tab:varDef} shows the definitions of the kinematic variables used in this study. 

\begin{longtable*}[H]{ll}
\caption{ Kinematic variables \label{tab:varDef} } \\ \hline \hline
$q=(E_{e}-E_{e}^{'},\vec{l}-\vec{l^{'}})$ & 4-momentum of the virtual photon \\
$Q^{2}=-q^{2}$	& Virtuality of the exchanged photon  \\
$P$ & 4-momentum of the proton \\
$E_{\gamma}$	& Energy of exchanged photon  \\
$x_{\gamma}$ & Momentum fraction of the parton from the exchanged photon\\
$x_{p}$ & Momentum fraction of the parton from the proton\\
$y=\frac{P\cdot q}{P\cdot l}$	& Energy fraction of virtual photon with respect to incoming electron \\
$\sqrt{s}$	& Center of mass energy \\
$p_{T}$ & Transverse momentum of final state particle(or jet) with respect to virtual photon \\
$\Delta\phi$ & Azimuthal angle difference of the two highest $p_{T}$ jets \\
$\eta =-\ln\tan(\theta/2)$ & pseudo-rapidity of the particles in lab frame \\
$\hat s$, $\hat t$, $\hat u$& Mandelstam variable for partonic processes \\ 

\hline \hline
\end{longtable*}

This article is organized as follows:
In Section~\ref{sec:detector} we briefly describe the detector requirement of tagging low $Q^{2}$ events.
We also discuss the framework used for measuring the structure of the photon.
The Monte Carlo simulations used to generate the di-jet cross section at a future EIC are validated by the data collected with the H1 detector at HERA.
In Section~\ref{sec:photonstructure} we present the method of distinguishing di-jets produced in resolved and direct processes,
and the measurement of di-jet cross sections in quasi-real photoproduction events in (un)polarized $ep$ collision is discussed.
Finally we close with a summary in Section~\ref{sec:summary}.

\section{Electron ion collider and simulation}\label{sec:detector}
\subsection{Low $Q^{2}$-tagger}\label{lowQ2tagger}

The eRHIC design~\cite{ECAschenauer} at BNL reuses the available infrastructure and facilities of RHIC's high-energy polarized proton and ion beams.
A new electron beam is to be built inside the current RHIC tunnel.
At eRHIC, the collision luminosity is expected to be of the order of $10^{33-34} \textrm{cm}^{-2}\textrm{s}^{-1}$.
The full range of proton/ion beam energies will be accessible from the beginning of operations,
with center-of-mass energies in the range 20 $\mathrm{GeV}$ to 140 $\mathrm{GeV}$.
A dedicated low $Q^{2}$-tagger is planned, to measure scattered electrons from low $Q^{2}$ events.
These electrons will miss the main detector, so installing an auxiliary device is essential for low $Q^{2}$ physics. 
Current designs for an EIC low $Q^{2}$-tagger
assume a lead tungstate (PbWO4) crystal calorimeter with a energy resolution of $2\%/\sqrt{E}+1\%$ preceded 
by Silicon detector planes for a high precision measurement of the  incident scattered electron angle.
The current design of the low $Q^{2}$-tagger essentially covers the region of 
$Q^{2}$ above $10^{-5}$ $\textrm{GeV}^{2}$. The present study is based on lepton and proton beam energies of $20$ 
$\textrm{GeV}$ $\times$ $250$ $\textrm{GeV}$, respectively.

\subsection{Monte Carlo Set Up}
 \begin{figure*} 
\begin{center}
\includegraphics[width=0.45\textwidth]{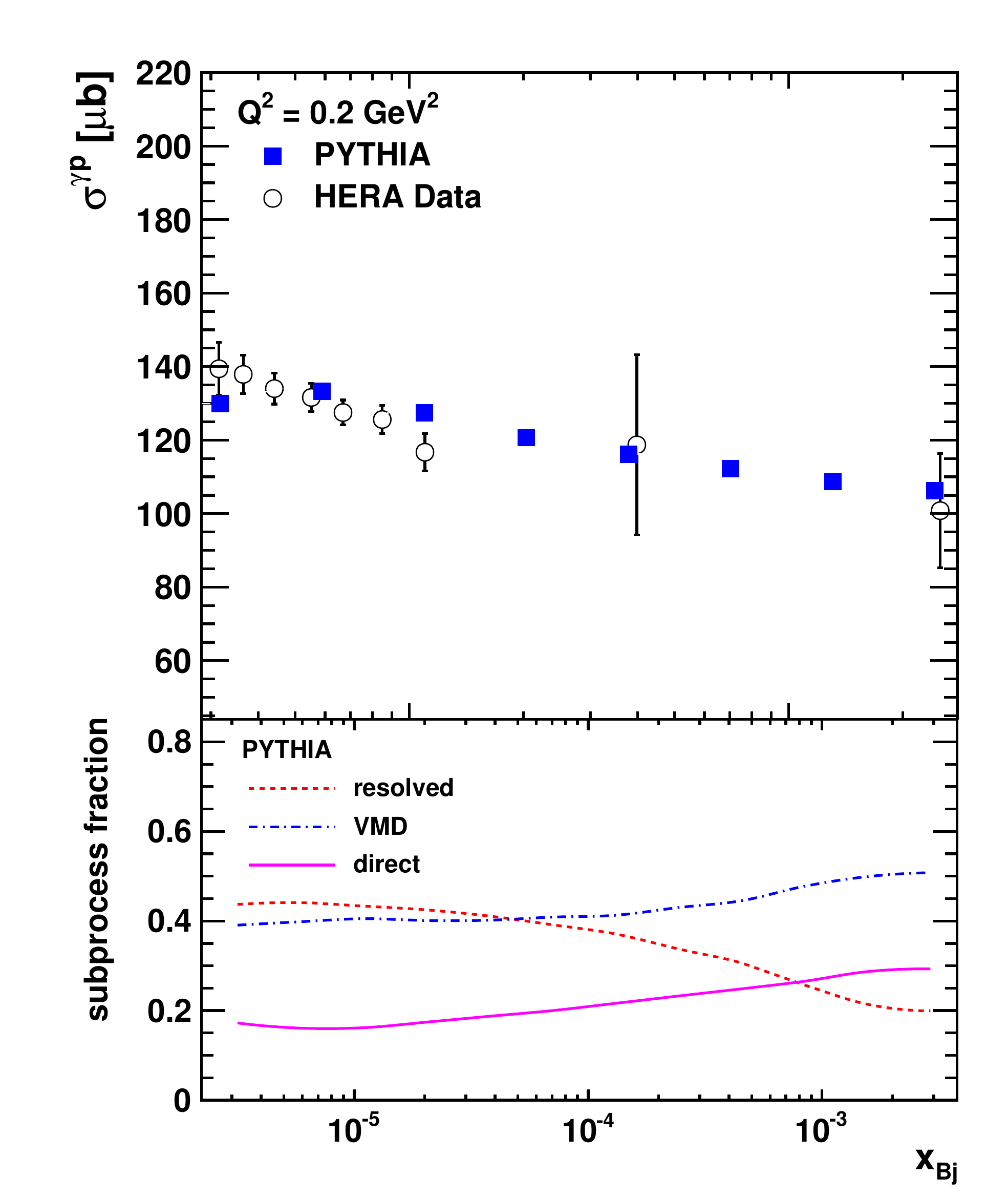}
\includegraphics[width=0.45\textwidth]{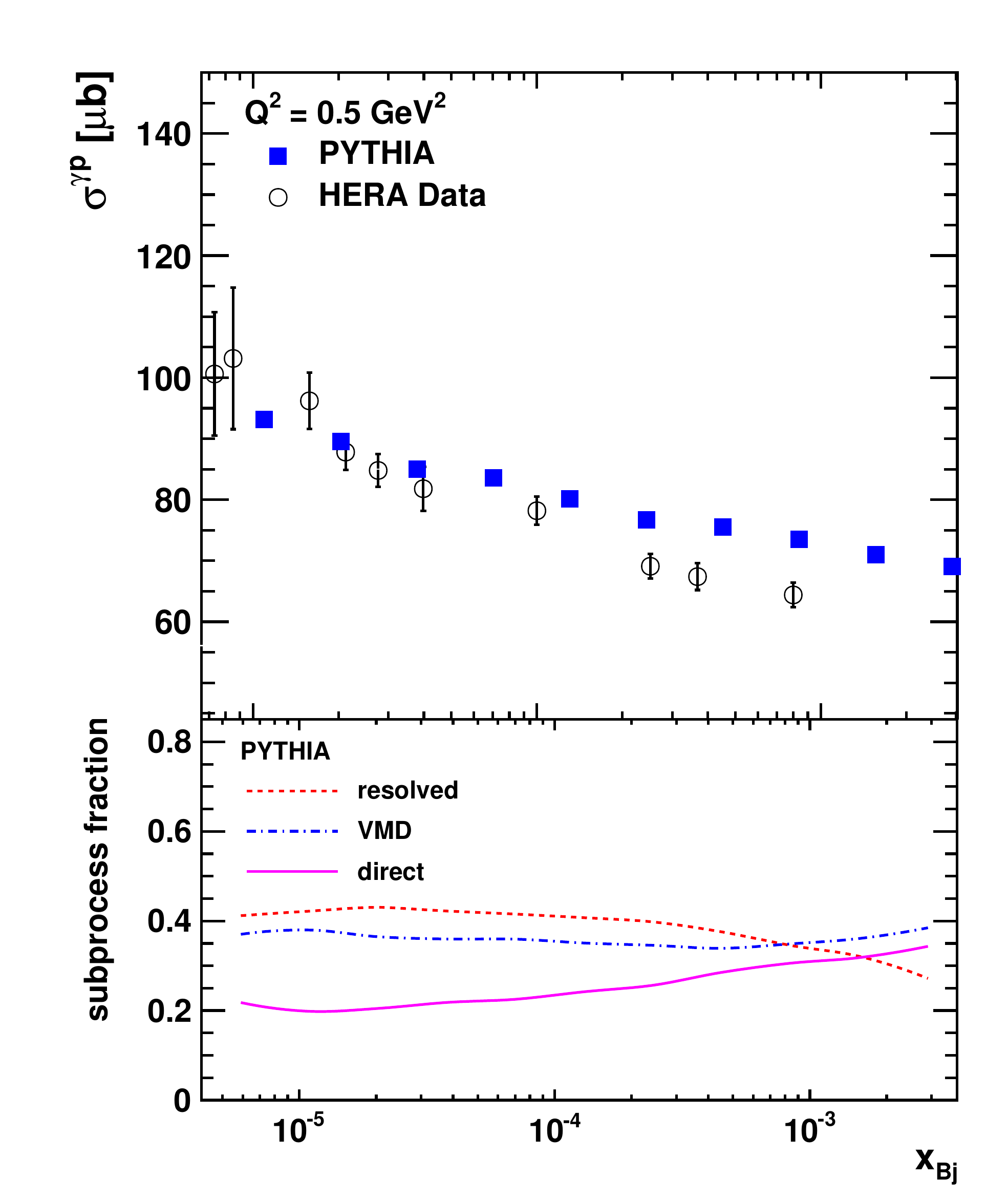}
\end{center} 
\caption{[color online] The $\sigma^{\gamma p}(x_{Bj}, Q^{2})$ simulated with PYTHIA-6 using
CTEQ5m and SAS 1D-LO as proton and photon PDFs, respectively, 
in comparison with the $\sigma^{\gamma p}(x_{Bj}, Q^{2})$ as extracted from the HERA $e^{+}p$ 
data~\cite{IAbt2}. 
Left: $Q^{2} =$ 0.2 GeV$^{2}$, Right: $Q^{2} =$ 0.5 GeV$^{2}$. 
Bottom: the different subprocess fractions for resolved, VMD and direct photon processes as a function of 
$x_{Bj}$.}
\label{fig:Inclusivexsection}
\end{figure*}

In this paper, we use pseudo-data generated by the Monte Carlo generator PYTHIA-$6$~\cite{Sjostrand},
with the unpolarized PDF input from the LHAPDF library~\cite{Whalley}.
In PYTHIA, depending on the wave function components for the incoming virtual
photon, the major hard processes are divided into three classes: the direct processes, the soft VMD processes
and the resolved processes (hard VMD and anomalous).
The direct photon interacts as a point-like particle with the partons of the nucleon, with major subprocesses in the direct category:
LO DIS, Photon-Gluon Fusion (PGF) and QCD Compton (QCDC).
The VMD and anomalous components interact through their hadronic structure.
Resolved photon processes play a significant part in the production of hard high-$p_{T}$ processes at 
$Q^{2}\approx0$. The following hard subprocesses are grouped in the resolved processes category: 
$qq\rightarrow qq, \, qg\rightarrow qg, \, gg\rightarrow gg, \, q\bar q\rightarrow q\bar q, \, gg\rightarrow q\bar q$ and $q\bar q\rightarrow gg$.
The CTEQ5m~\cite{CTEQ5} PDF is used for the proton, because
contrary to modern PDFs (i.e., CT, NNPDF, HERAPDF, MSTW) its PDF is not frozen at its input scale 
$Q^{2}_{0}$, 
but allows description of the partonic structure of the proton at $Q^2 \leq Q^{2}_{0}$.

The simulation used SAS 1D-LO~\cite{SAS} as photon-PDF.
This was for several reasons.
Most currently existing photon PDFs (DG-G~\cite{DG}, LAC-G/GAL-G~\cite{LAC}, GS-G~\cite{GS}, GS-G-96~\cite{GS96}, GRV-G/GRS-G~\cite{GRV}, ACFGP/AFG-G~\cite{ACFGP}, WHIT-G~\cite{SAS1} and SAS-G(v1/v2)~\cite{SAS1})
are only constrained by fits to the sparse $F_{2}$ data from electron-positron colliders~\cite{F2} before LEP;
the older DO-G~\cite{DO} PDF is based on low energy photon-proton data.
None of the existing Photon PDFs has HERA H1~\cite{H1} or ZEUS~\cite{ZEUS} data 
sensitive to the partonic structure of the photon included in the fits. 
References \cite{H1Collaboration5,H1,H11} discuss that the H1 data are best described by the SAS and 
GRV Photon PDFs. But as none of the Photon PDFs provide an evaluation of an uncertainty band 
as is standard for the current Proton PDFs, and with the statistical precision of the HERA data remaining 
limited, no real quantitative preference for any of the photon PDF sets can be determined.
The SAS PDF is best suited for use in PYTHIA, since the vector meson and anomalous photon components
are unfolded, thus avoiding double counting of resolved photon subprocesses. 
FIG.~\ref{fig:Inclusivexsection} (upper plots) shows an excellent agreement between the PYTHIA-6 
simulation using CTEQ5m as the Proton PDF and SAS 1D-LO as 
the Photon PDF and the low $Q^{2}$ data from HERA~\cite{IAbt2}. The lower parts of 
FIG.~\ref{fig:Inclusivexsection} show the mix of direct and resolved (hard VMD and anomalous) photon 
processes.

FastJet~\cite{MCacciari} is used for jet reconstruction.
The kinematics are constrained to the region of interest for photoproduction by requiring that
the scattered electron remains in the beam pipe, undetected in the main detector.
The photon virtuality is therefore restricted to $Q^{2}<0.1$ $\textrm{GeV}^{2}$ and $Q^{2}>10^{-5}$ $\textrm{GeV}^{2}$,
according to the lower limit of the low $Q^{2}$-tagger acceptance. 

\subsection{Verification of Simulation with HERA data}

Predictions for the di-jet cross section in photoproduction events are obtained in leading order quantum chromodynamics (LO QCD)
by convoluting the parton densities in the photon and those in the proton with the short-distance partonic cross section,
\begin{equation}\label{xgammarec}
\frac{d^{2}\sigma}{dx_{\gamma}dQ^{2}}= \gamma_{flux}\otimes f_{\gamma}(x_{\gamma},Q^{2},\mu)\\ 
\otimes f_{p}(x_{p},\mu) \otimes \sigma_{ij},
\end{equation}
where $\gamma_{flux}$ is the flux of photons emitted from the incoming electron.
The fractional momentum of the parton in the photon is given by $x_{\gamma}$ and parton density function of the photon by $f_{\gamma}$.
The corresponding variables for the proton are $x_{p}$ and $f_{p}$.
$\sigma_{ij}$ is the hard cross section of the subprocess. 
Assuming the parton densities in the proton are well known, a measurement of the di-jet cross section
can be used to extract information on the parton densities of the photon.
In order to extract the photon structure information in di-jet production, we need to distinguish resolved processes from direct processes.
The best variable to separate the two types of processes is $x_{\gamma}$.
Since in the direct processes the photon interacts with a parton from the proton as a structureless particle with its whole energy entering the hard scattering,
the $x_{\gamma}$ of direct processes is equal to 1.
In the resolved processes, the photon behaves like a source of partons, with only a fraction of its momentum participating in the hard scattering;
therefore the corresponding $x_{\gamma}$ should be smaller than 1. 

\begin{figure}[hbt]
 \centering
  \includegraphics[width=0.49\textwidth]{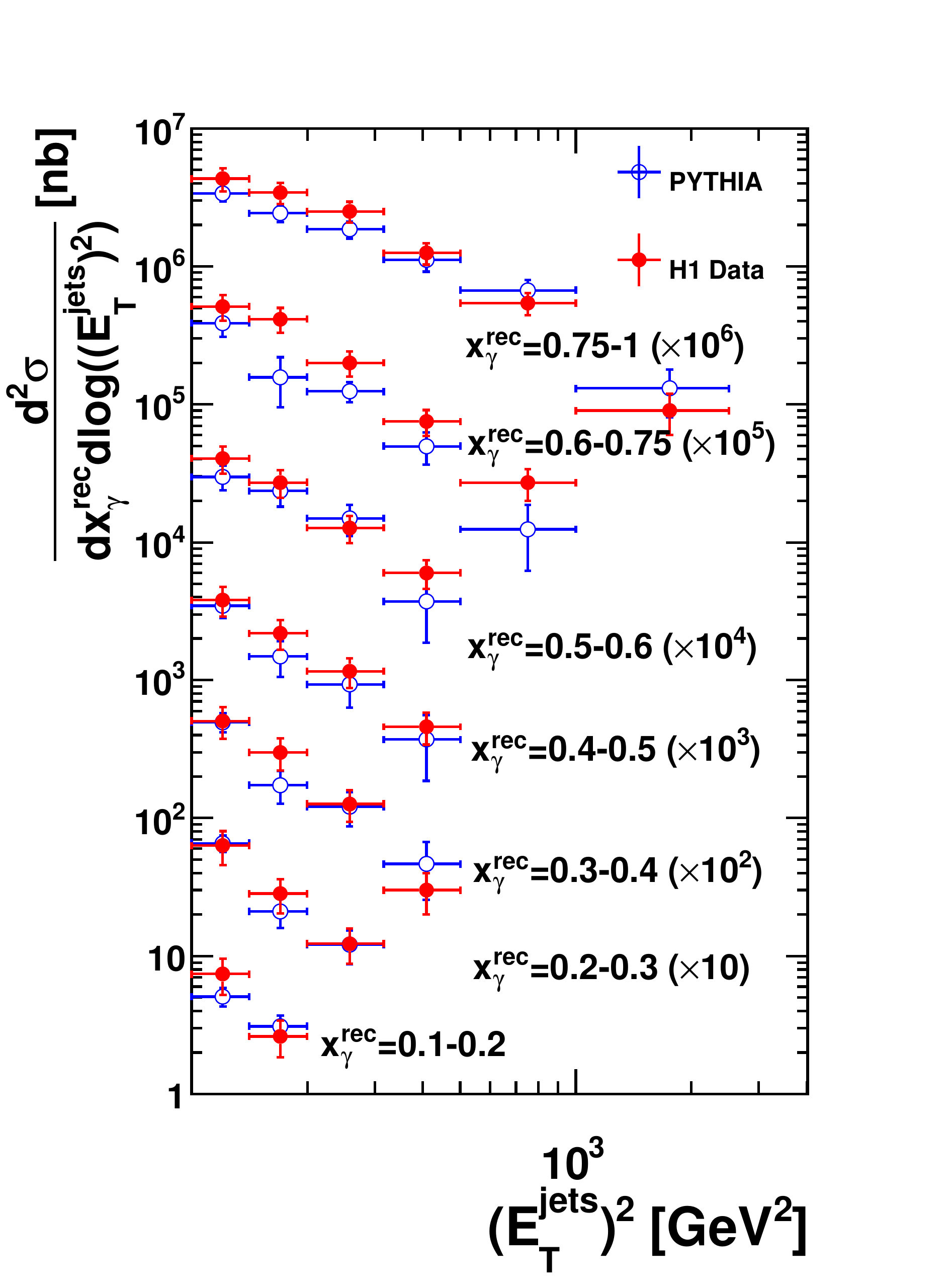}
\caption{[color online] Comparison of the di-jet cross section extracted from the PYTHIA simulation 
with the HERA data.  The kinematics cuts are from HERA:
$E_{T}^\textrm{jets}>10$ $\mathrm{GeV}$, $\frac{|E_{T}^\textrm{jet1}-E_{T}^\textrm{jet2}|}{(E_{T}^\textrm{jet1}+E_{T}^\textrm{jet2})}<0.25$,
the photon virtuality $Q^{2}< 4$ $\mathrm{GeV^{2}}$, the fractional photon energy is between 
$0.2<y<0.83$, and the average of pseudo-rapidity of the two jets is restricted to 
$0 < \frac{\eta^\textrm{jet1}+\eta^\textrm{jet2}}{2} <2$ and $|\Delta\eta^\textrm{jets}| < 1$.
The H1 data is from~\cite{H1Collaboration6}.}
\label{fig:HERAcomparison}
\end{figure}

The variable $x_{\gamma}$ can be reconstructed from the momenta and angles of di-jets as
\begin{equation}\label{eqa:xrec}
x_{\gamma} =  \frac{1}{2E_{e}y} (p_{T,1}e^{-\eta_{1}}+p_{T,2}e^{-\eta_{2}}) ,
\end{equation}
where $E_{e}$ is the electron beam energy and $y$ is the energy fraction taken by the photon from the electron ($y=\frac{E_{\gamma}}{E_{e}}$).
Eq.~\ref{eqa:xrec} is valid in the lab frame in LO.

The di-jet cross section measured by H1 at HERA~\cite{H1Collaboration6} is shown in FIG.~\ref{fig:HERAcomparison}
as a function of the squared jet transverse energy $E_{T}^\textrm{jets}$ in ranges of reconstructed $x_{\gamma}$.
Here $E_{T}^\textrm{jets}$ is the average transverse energy of the two highest $p_{T}$ jets:
($E_{T}^\textrm{jets}=\frac{E_{T}^\textrm{jet1}+E_{T}^\textrm{jet2}}{2}$).
$E_{T}^\textrm{jets}$ is required to be above $10$ $ \mathrm{GeV}$.
The ratio of the difference and the sum of the transverse energies of the jets is required to satisfy
$\frac{|E_{T}^\textrm{jet1}-E_{T}^\textrm{jet2}|}{(E_{T}^\textrm{jet1}+E_{T}^\textrm{jet2})}<0.25$,
and the transverse energy of individual jets is required to be above $7.5$ $ \mathrm{GeV}$.
The fractional photon energy is restricted to $0.2<y<0.83$.
The average of the pseudo-rapidity of the two jets is restricted to $0 < \frac{\eta^\textrm{jet1}+\eta^\textrm{jet2}}{2} <2$,
and the difference of the jet pseudo-rapidities is required to be within $|\Delta\eta^\textrm{jets}| < 1$.
The simulation results are obtained for 27 $\mathrm{GeV}$ electrons colliding with protons of 820 $\mathrm{GeV}$,
and the comparison of our simulation with the H1 data shows that the simulation reproduces the measured data well.
Some of the observed difference is due to the use of the anti-$k_{T}$ algorithm~\cite{MCacciari1} for the jet finding
instead of the cone algorithm used for the HERA results. 

\section{Photon structure at EIC}\label{sec:photonstructure}
\subsection{The Unpolarized Photon Structure }\label{sec:UnpolarizedPhotonMC}

\begin{figure}[hbt]
\centering
  \includegraphics[width=0.45\textwidth]{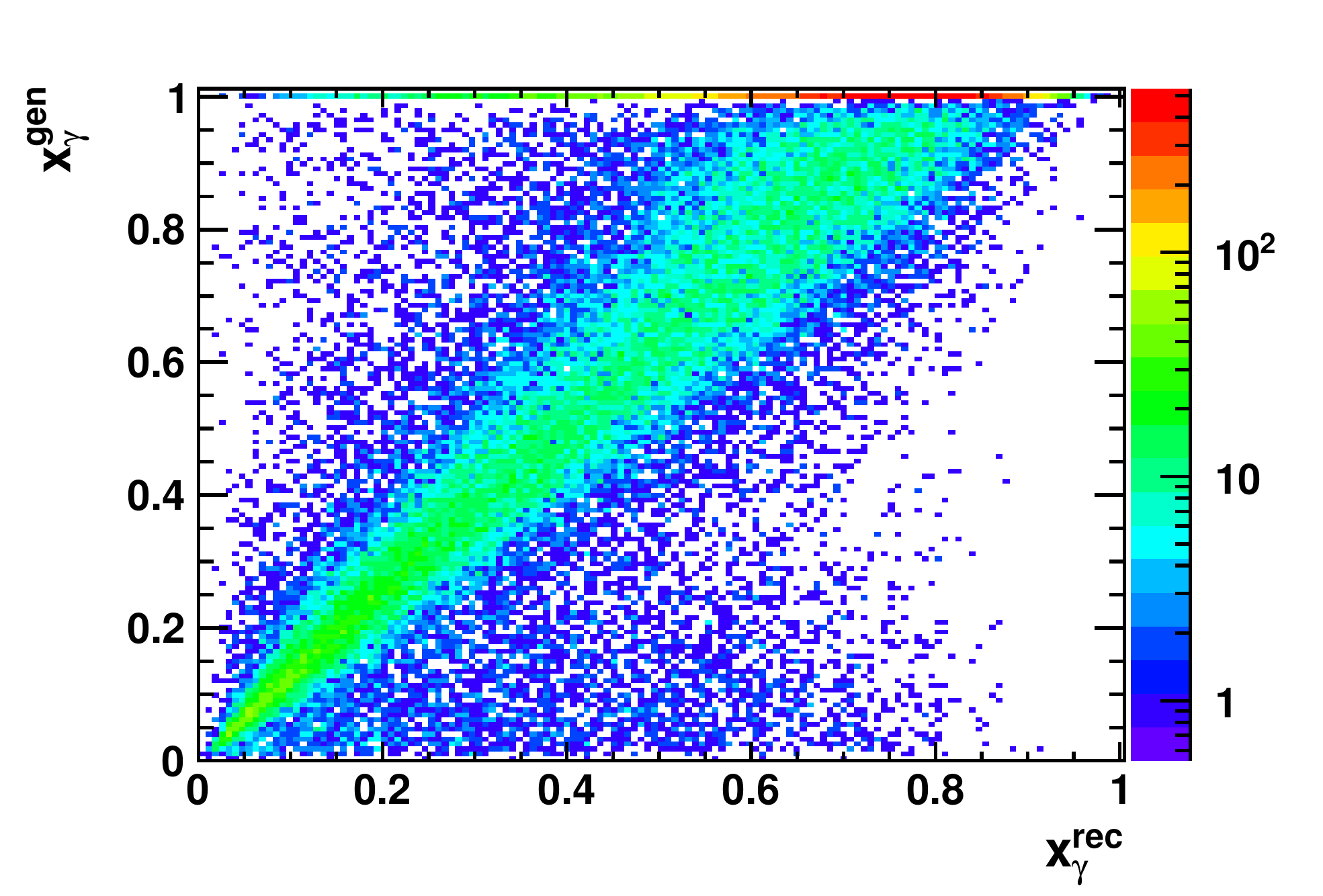}
  \caption{[color online] Correlation between $x_{\gamma}^{gen}$ and $x_{\gamma}^{rec}$.
  A di-jet is reconstructed in each event with stable particles, the particles inside the jets require to be $p_{T}>250$ $\textrm{MeV/c}$ and $-4.5<\eta<4.5$.
  The di-jet events are selected with $10^{-5}$ $\mathrm{GeV^{2}}<Q^{2}< 0.1$ $\mathrm{GeV^{2}}$, $N^\textrm{jets}\ge 2$,
  trigger jet $p_{T}^\textrm{jet1}> 5$ $\mathrm{GeV/c}$, and associated jet $p_{T}^\textrm{jet2}> 4.5$ $\mathrm{GeV/c}$.}
\label{fig:xrec_xbparton}
\end{figure}

\begin{figure}[hbt]
 \centering
  \includegraphics[width=0.45\textwidth]{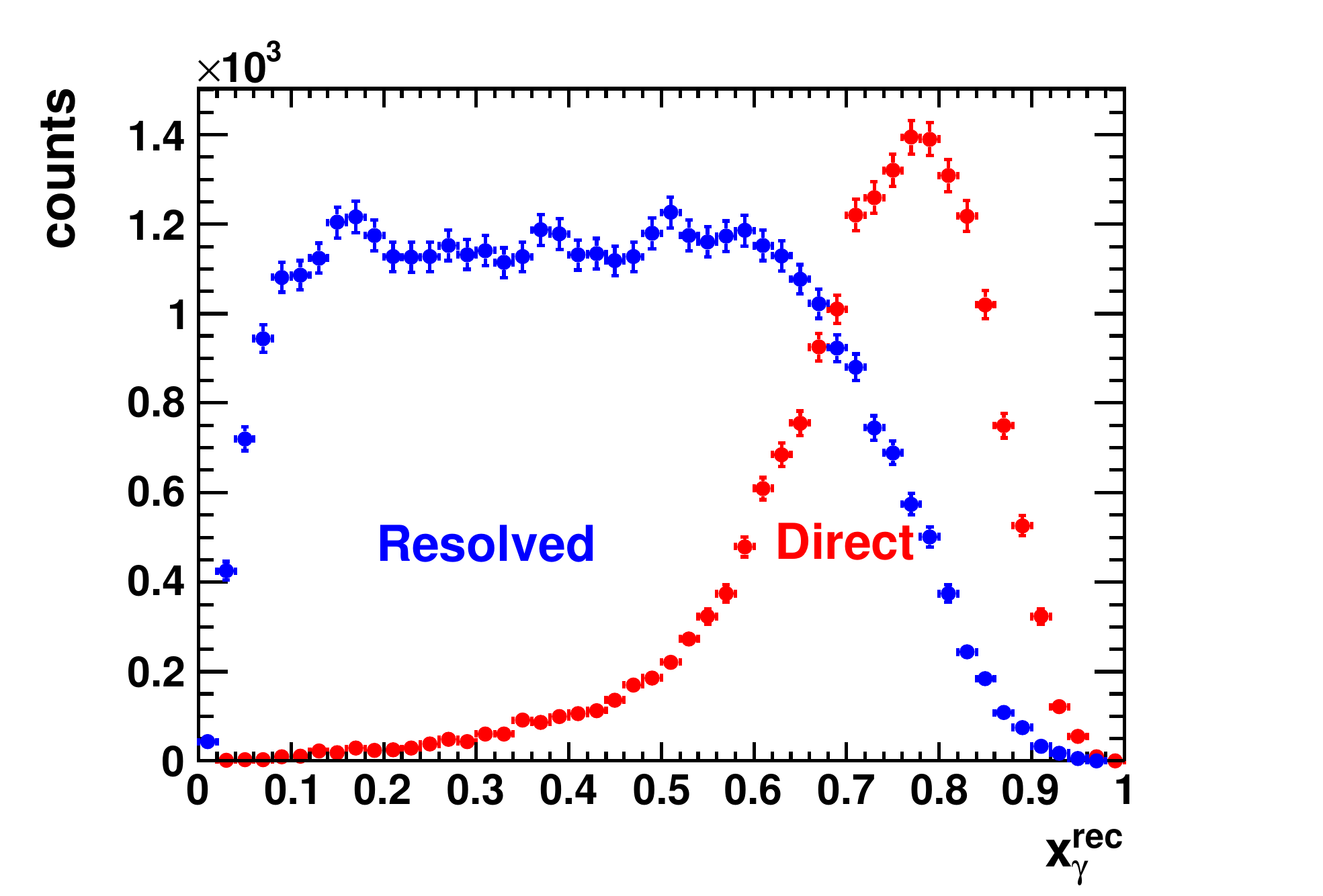}
  \caption{[color online] $x_{\gamma}^{rec}$ distributions in resolved and direct processes.}
\label{fig:xrec}
\end{figure}

\begin{figure}[hbt]
  \centering
  \includegraphics[width=0.49\textwidth]{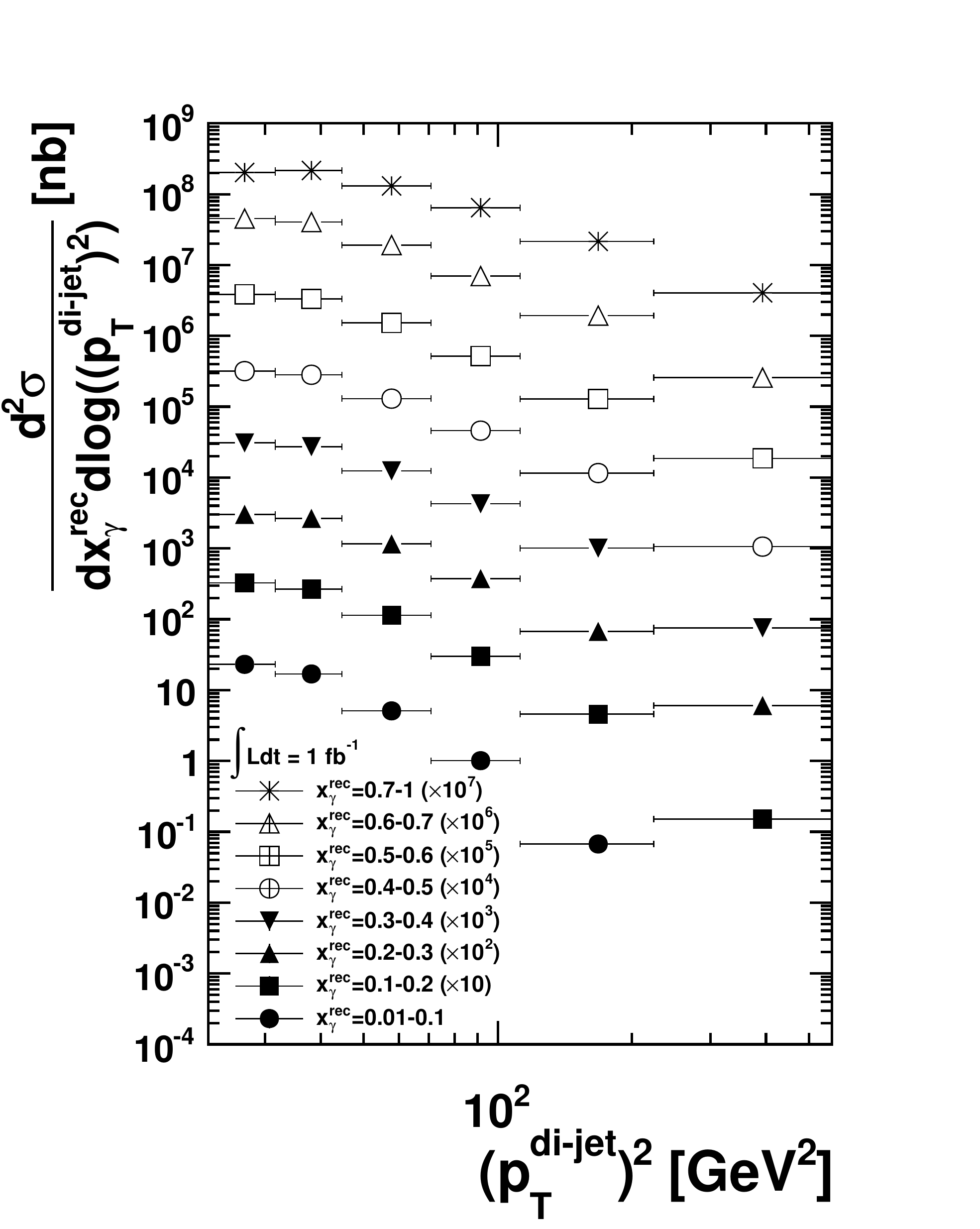}
  \caption{The unpolarized di-jet cross section dependence on the average transverse momentum of the jets
  $p_{T}^\textrm{di-jet}= \frac{p_{T}^\textrm{jet1}+p_{T}^\textrm{jet2}}{2}$ and the reconstructed $x_{\gamma}$ for an integrated luminosity of 1 $fb^{-1}$.
  Low $Q^{2}$ events are selected: $10^{-5}$ $\mathrm{GeV^{2}}<Q^{2}< 0.1$ $\mathrm{GeV^{2}}$, $N^\textrm{jets}\ge 2$.
  The anti-$k_{T}$ algorithm is used with $R=1$. For the trigger jet $p_{T}^\textrm{jet1}> 5$ $\mathrm{GeV/c}$,
  and the associated jet $p_{T}^\textrm{jet2}> 4.5$ $\mathrm{GeV/c}$.}
  \label{fig:unxsection}
\end{figure}

In this analysis, jets are reconstructed with the anti-$k_{T}$ algorithm, which is based on the energy distribution of final state particles in the angular space.
All the stable and visible particles produced in the collision with $p_{T}>250$ MeV$/c$ and $-4.5<\eta<4.5$ in the laboratory system are taken as input.
The jet cone radius parameter has been set to $R=1$ in the jet finding algorithm.

\begin{figure*} 
\begin{center}
\includegraphics[width=0.45\textwidth]{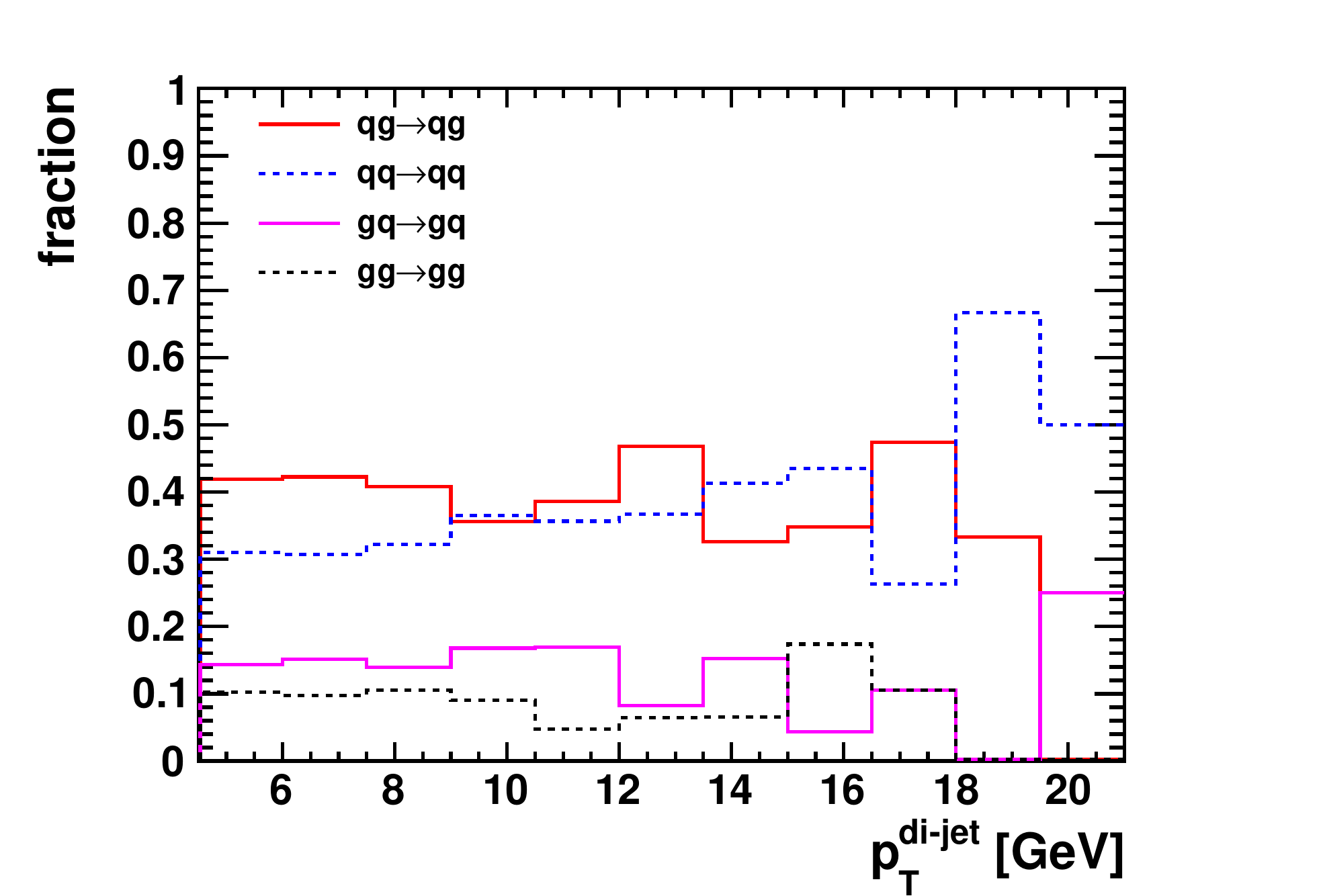}
\includegraphics[width=0.45\textwidth]{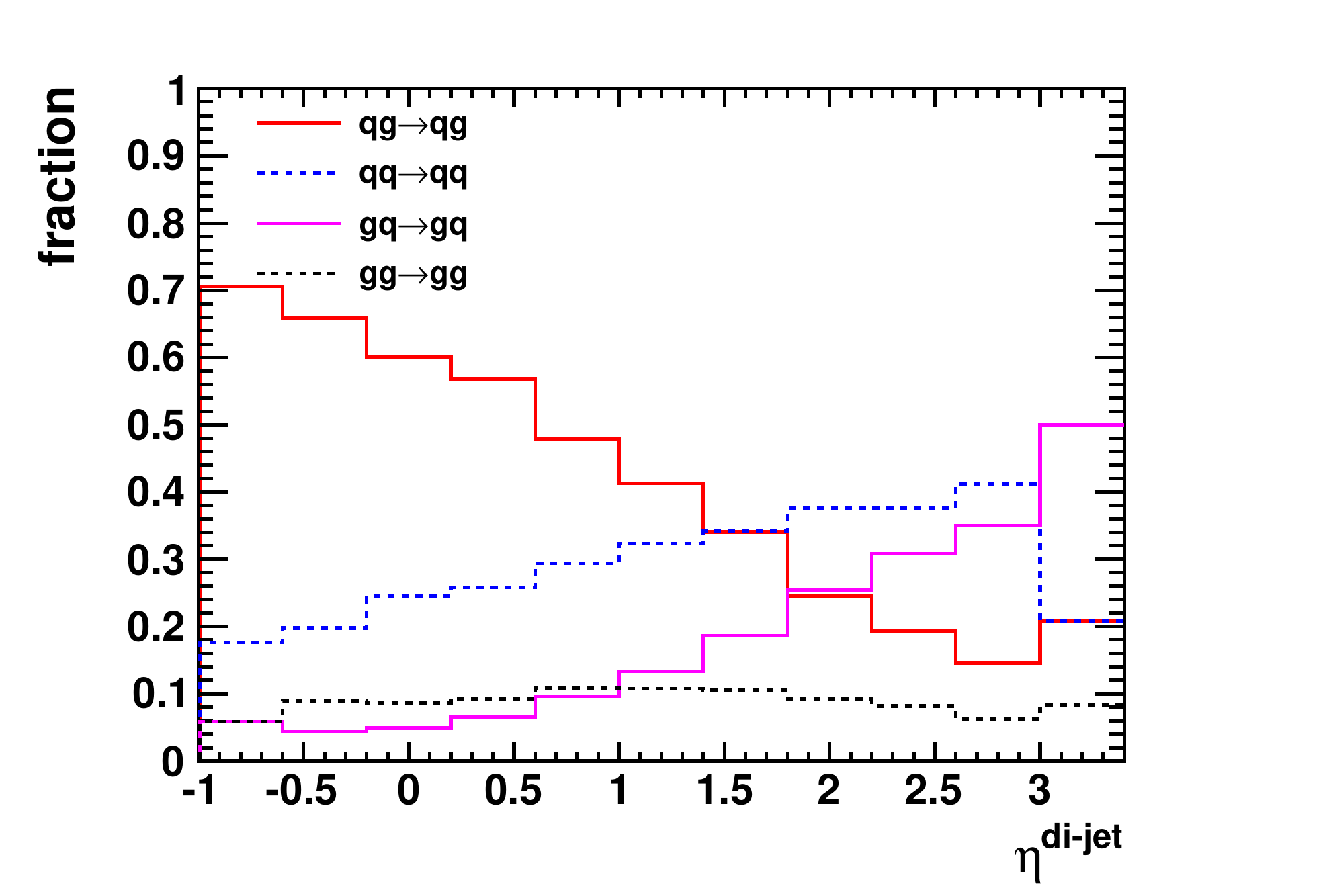}
\end{center} 
\caption{[color online] Left: the fraction of the major subprocesses of the resolved process dependence on $p_{T}^\textrm{di-jet}$.
Right: the fraction distribution dependence on $\eta^\textrm{di-jet} = \frac{\eta^\textrm{jet1}+\eta^\textrm{jet2}}{2}$.}
\label{fig:Sub}
\end{figure*}

This simulation is performed for the planned EIC electron and proton beam energy configuration of 20 $\mathrm{GeV}$ $\times$ 250 $\mathrm{GeV}$.
We consider events with two or more jets; 85\% of all events have exactly two jets ($p_{T}^\mathrm{jet}>3$ $\mathrm{GeV/c}$).
In each event, the jet with the highest $p_{T}$ is referred to as the trigger jet, and the jet with the second highest $p_{T}$ the associated jet.
Events are selected with the requirement that the trigger jet has $p_{T}^\textrm{jet}> 5$ $\mathrm{GeV/c}$,
the associated jet has $p_{T}^\textrm{jet}> 4.5$ $\mathrm{GeV/c}$, and $10^{-5}$ $\mathrm{GeV^{2}}<Q^{2} < 0.1$ $\mathrm{GeV^{2}}$.
The average transverse momentum of the trigger and associated jets is $p_{T}^\textrm{di-jet}= \frac{p_{T}^\textrm{jet1}+p_{T}^\textrm{jet2}}{2}$.
In this analysis, the event kinematic variables $Q^{2}$ and $y$ are obtained directly from 
PYTHIA simulations without reconstructing them from the event information.
The variable $y = E_{\gamma}/E_{e}$ can be experimentally reconstructed in two ways.
The scattered lepton, if detected in the low $Q^{2}$-tagger as described in subsection~\ref{lowQ2tagger},
provides a direct measurement of $y$.
The other possibility to reconstruct $y$ is through the Jacquet-Blondel method~\cite{Jacquet}, which 
utilizes the hadronic final state. Reference~\cite{elke} discussed this method and its performance for charged 
current events at an EIC.

We reconstruct $x_{\gamma}$ in di-jet events according to Eq.~{\ref{eqa:xrec}}.
The strong correlation between the reconstructed $x_{\gamma}^{rec}$ and the true $x_{\gamma}^{gen}$ in PYTHIA is shown in FIG.~\ref{fig:xrec_xbparton}.
It clearly shows that the di-jet observable is ideal for this measurement.
The $x_{\gamma}^{rec}$ distribution for the resolved (direct) process dominates in the low (high) $x_{\gamma}^{rec}$ regime (see FIG.~\ref{fig:xrec}),
which provides good separation of the two types of processes.
For example, by selecting events with $x_{\gamma}^{rec} \sim 0$ or $x_{\gamma}^{rec} \sim 1$, one can divide the di-jet events into subsamples
in which the resolved and direct processes dominate, respectively.
As a smaller $x_{\gamma}^{rec}$ cut is chosen higher purity for the resolved process is obtained.
Considering the balance between statistics and purity, $x_{\gamma}^{rec} < 0.6$ is chosen;
with this cut the fraction of the resolved process ($N_\textrm{res}/(N_\textrm{res} + N_\textrm{dir})$) is up to $91.2\%$.

FIG.~\ref{fig:unxsection} shows the resulting high precision double differential di-jet cross section over a wide kinematic range
with an integrated luminosity of $1 fb^{-1}$.
With a global fit the unpolarized photon PDFs can be extracted from the cross section. 

\subsection{Flavor Tagging}\label{sec:FlavorTagging}

The resolved process has several types of subprocesses, divided into $7$ types:
$qq\rightarrow qq$ (which means $q_{i}q_{j}\rightarrow q_{i}q_{j}$, with $q_{i}q_{j}$ standing for both quark or anti-quark),
$q$(photon)$g\rightarrow q$(photon)$g, \, g$(photon)$q\rightarrow g$(photon)$q, \, gg\rightarrow gg$, $q\bar q\rightarrow q\bar q$ (which means $q_{i}\bar q_{i}\rightarrow q_{k}\bar q_{k}$), $gg\rightarrow q\bar q$ and $q\bar q\rightarrow gg$.
Since the first four types of subprocess account for more than $96\%$ percent of the resolved process, we  mainly discuss these four types.
The subprocess fraction depends on the average transverse momentum of the di-jet, as shown in the left of FIG.~\ref{fig:Sub}.
The process $qq\rightarrow qq$ is more likely to dominate in the large $p_{T}^\textrm{di-jet}$ region.
Gluon jets produced in $gg\rightarrow gg$ process are softer.
As shown in the right of FIG.~\ref{fig:Sub}, the fraction of the different subprocesses depends on the average pseudo-rapidity.
The $qg\rightarrow qg$ process dominates in the negative $\eta^\textrm{di-jet}$ region.

\begin{figure*} 
\begin{center}
\includegraphics[width=0.45\textwidth]{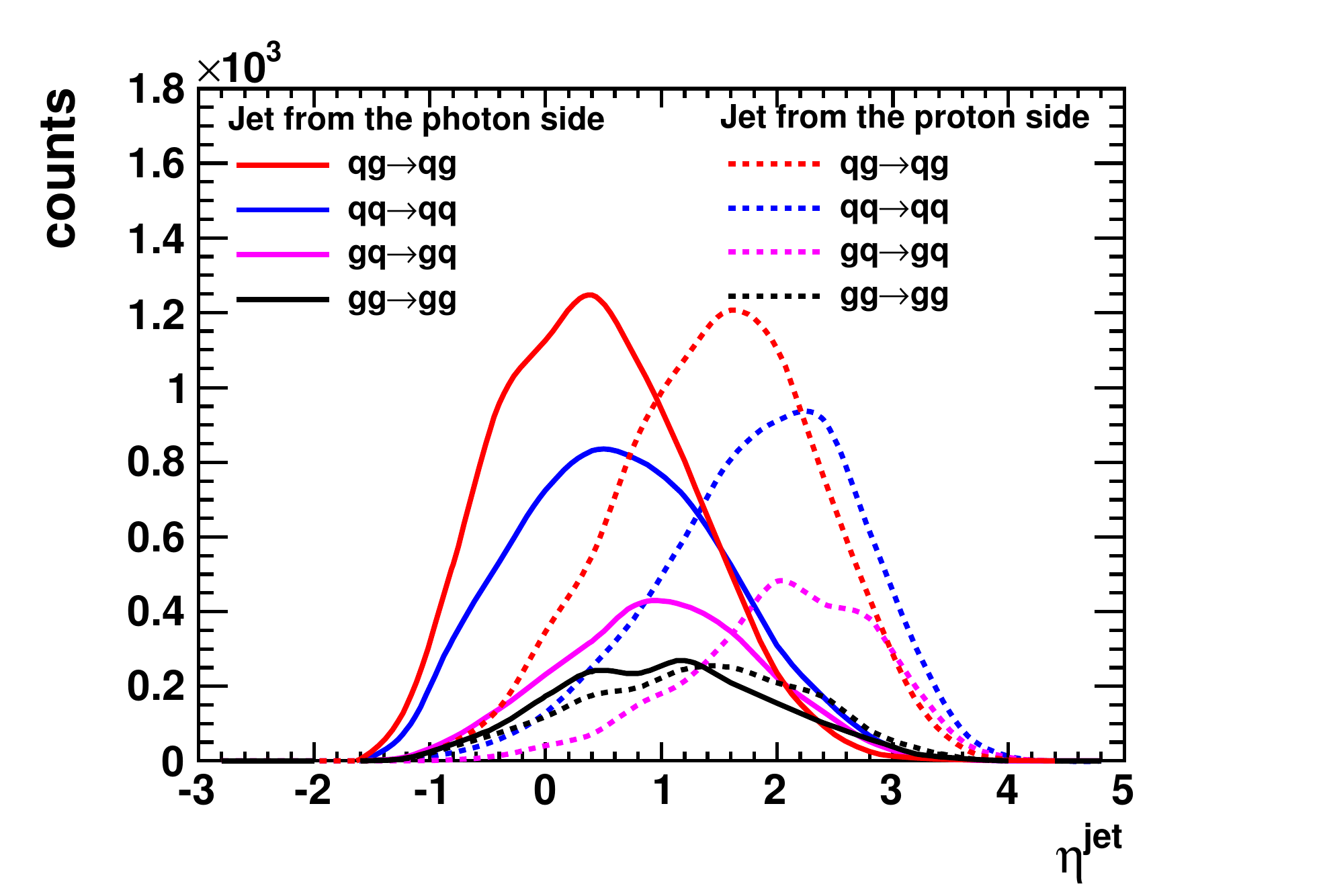}
\includegraphics[width=0.45\textwidth]{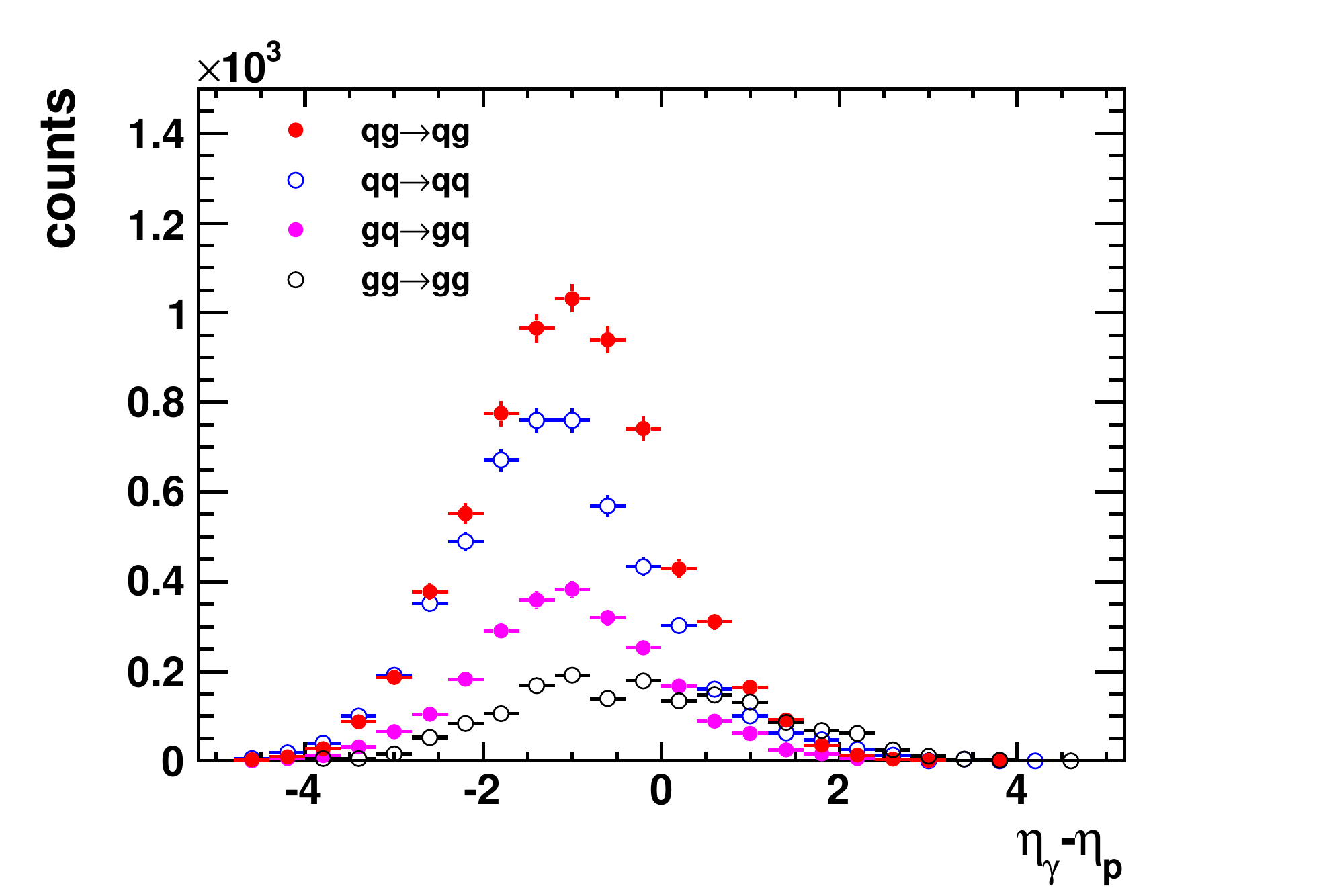}
\end{center} 
\caption{[color online] Left: the pseudo-rapidity distribution of jets from the photon side and proton side in different subprocesses.
Right: $\Delta\eta$ distribution between jets from the photon side and jets from the proton side on an event-by-event analysis.}
\label{fig:GammaPJetSepa}
\end{figure*}

In order to precisely determine the photon PDF for different parton components,
it is important to devise an experimental handle on the flavor of the parton involved in the hard 
interaction from the photon side. 
Tagging the parton flavor through identified hadrons in jets has recently become an 
important tool, especially in $pp$ collisions,  to study PDFs and fragmentation functions.
(For theoretical details and 
first experimental results see Reference~\cite{TMDs, FFs} and \cite{ATLASFFs,STAR}). 
In the following, we apply this method to tag the parton flavor content of the photon at an EIC.

We demonstrate in this section that the outgoing jet close to the electron beam pseudo-rapidity 
is more likely to take the incoming parton flavor from the photon side.
The leading hadron species inside those jets are found to be strongly correlated to the underlying 
parton flavors.
Then a straightforward strategy is to tag the parton flavor of the photon through the leading hadron 
type inside the photon side jet.
We find a high cut on the transverse momentum fraction of the leading hadron will enhance the 
sensitivity to the parton flavor even further.

In the following we call the parton coming from the photon the ``beam parton" and the one coming from the proton the ``target parton".
In a leading order $2\rightarrow2$ scattering process in quasi-real photoproduction events, the beam parton is converted to the jet from the photon side,
while the target parton to the jet from the proton side.
In our simulation, it is possible to apply a geometric match between the outgoing partons and the jets.
Therefore, in each di-jet event, the jet from the photon side and the jet from the proton side are accessible in the simulation.
We find in our simulation that the pseudo-rapidity distribution of the outgoing jets from the photon and proton sides are distinguishable,
as shown in the left of FIG.~\ref{fig:GammaPJetSepa}; an exception is that gluon jets from both sides overlap with each other in the $gg\rightarrow gg$ process.
If we define the proton beam direction as the positive pseudo-rapidity direction,
jets from the photon side dominate at more negative pseudo-rapidities compared with jets from the proton side.

On an event-by-event basis, the pseudo-rapidity difference between the two outgoing jets $\Delta\eta = \eta^{\textrm{jet}}_{\gamma}-\eta^{\textrm{jet}}_{p}$
can be used to identify the photon and proton side jets.
As shown in the right of FIG.~\ref{fig:GammaPJetSepa}, the value of $\Delta\eta$ is mostly negative.
The $gg\rightarrow gg$ process is the only exception, so the quark jet from the photon side can be well identified.
For 82.0\% of the events the jets from the photon side take the more negative pseudo-rapidity than those from the proton side,
which provides an experimental way to separate jets from the photon side and jets from the proton side:
in each event, we take the jet with more negative pseudo-rapidity as the jet from the photon.

\begin{figure*}
\begin{center}
\includegraphics[width=0.8\textwidth]{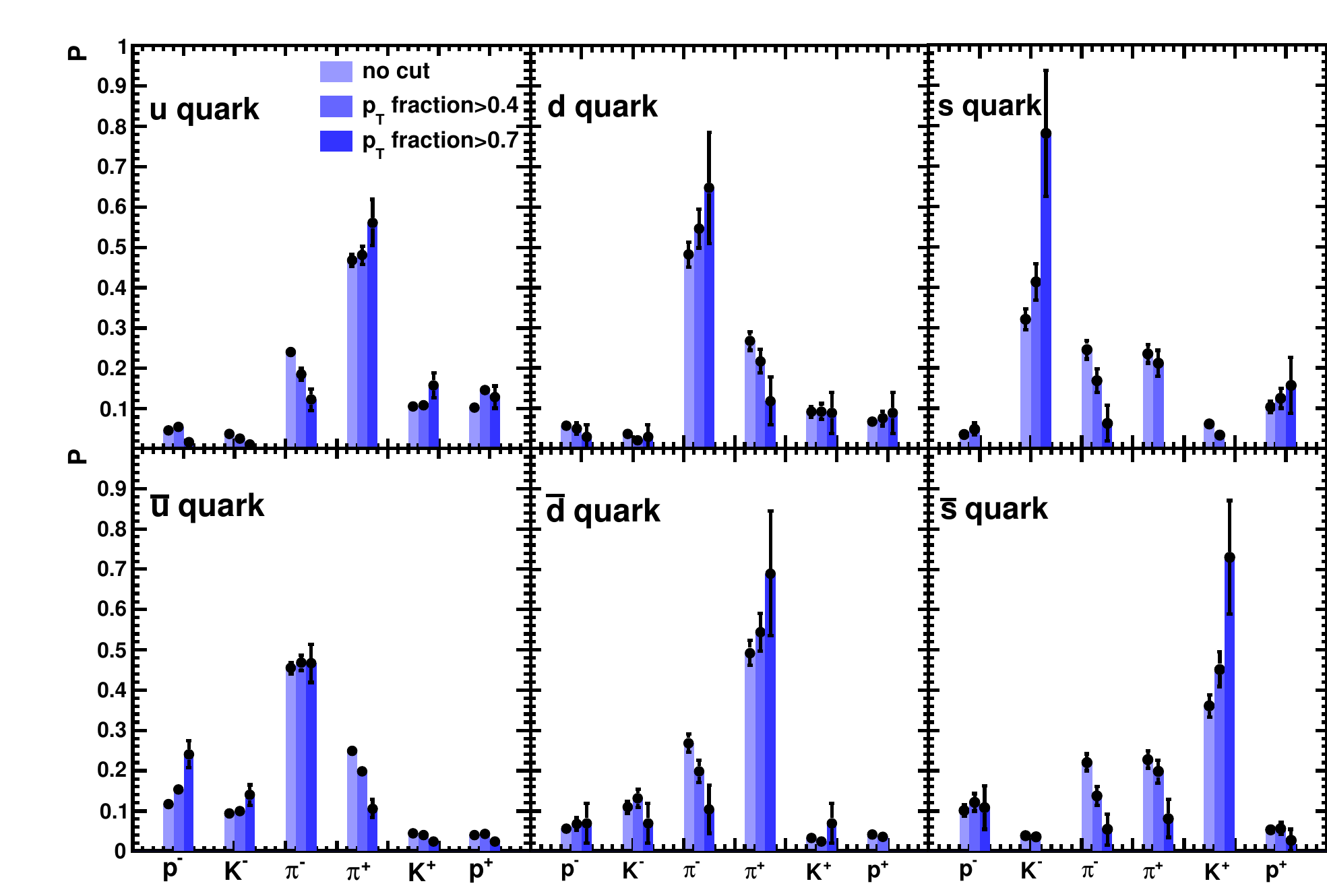} 
\end{center} 
\caption{[color online] Probability P of finding a leading charged hadron depending on different parton flavors of the photon
for three different cuts on the $p_{T}$ fraction.
$p_{T}$ fraction is measured as $\frac{p_{T}^{L}}{p_{T}^\textrm{{jet}}}$.}
\label{fig:flavortagging} 
\end{figure*}
\begin{figure*}
\begin{center}
\includegraphics[width=0.7\textwidth]{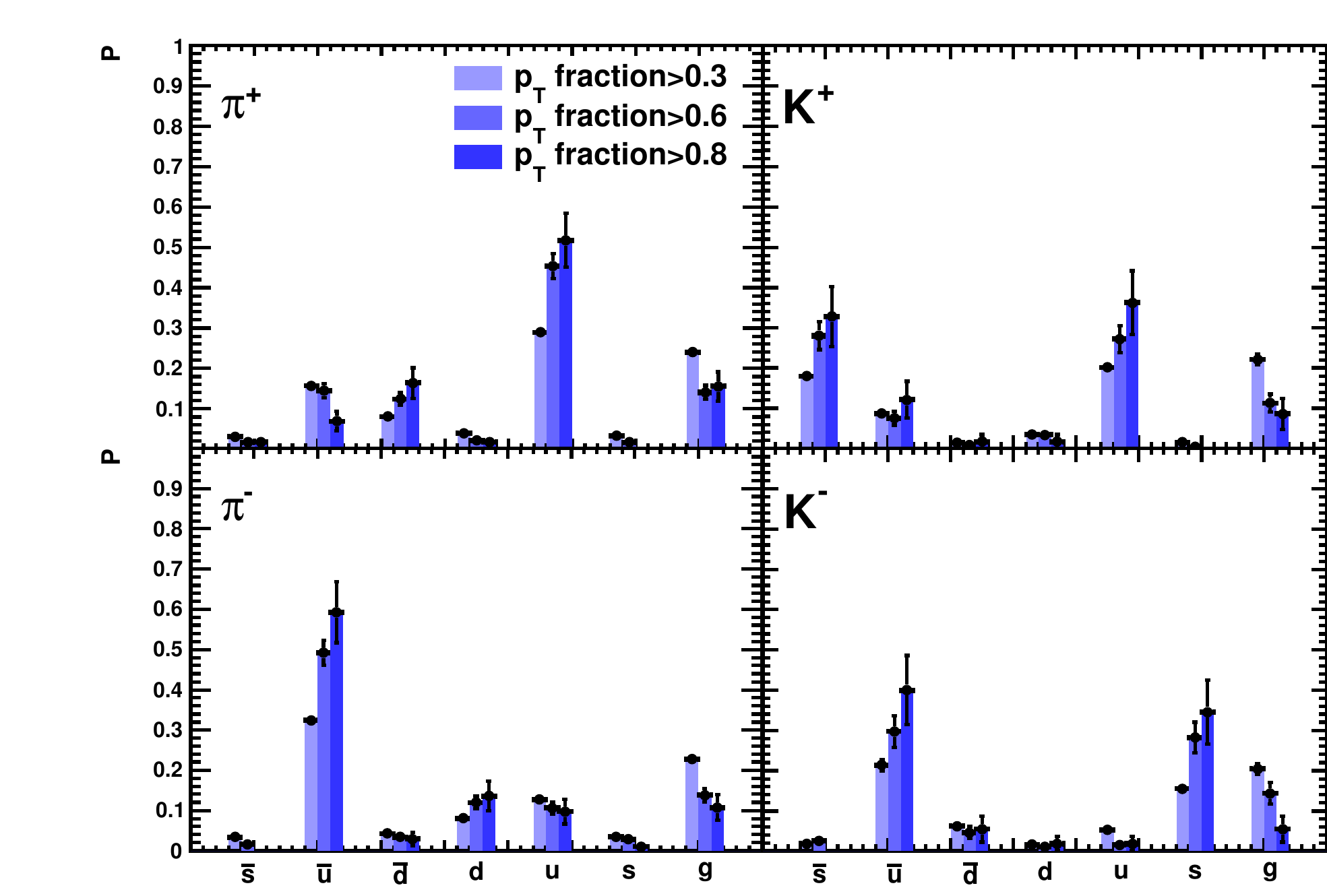} 
\end{center} 
\caption{[color online] Probability P of the parton flavor in the photon jet depending on the leading hadron type in the photon side jet
for three different cuts on the $p_{T}$ fraction.
}
\label{fig:PKbeam parton} 
\end{figure*}

\begin{figure*} 
\begin{center}
\includegraphics[width=0.45\textwidth]{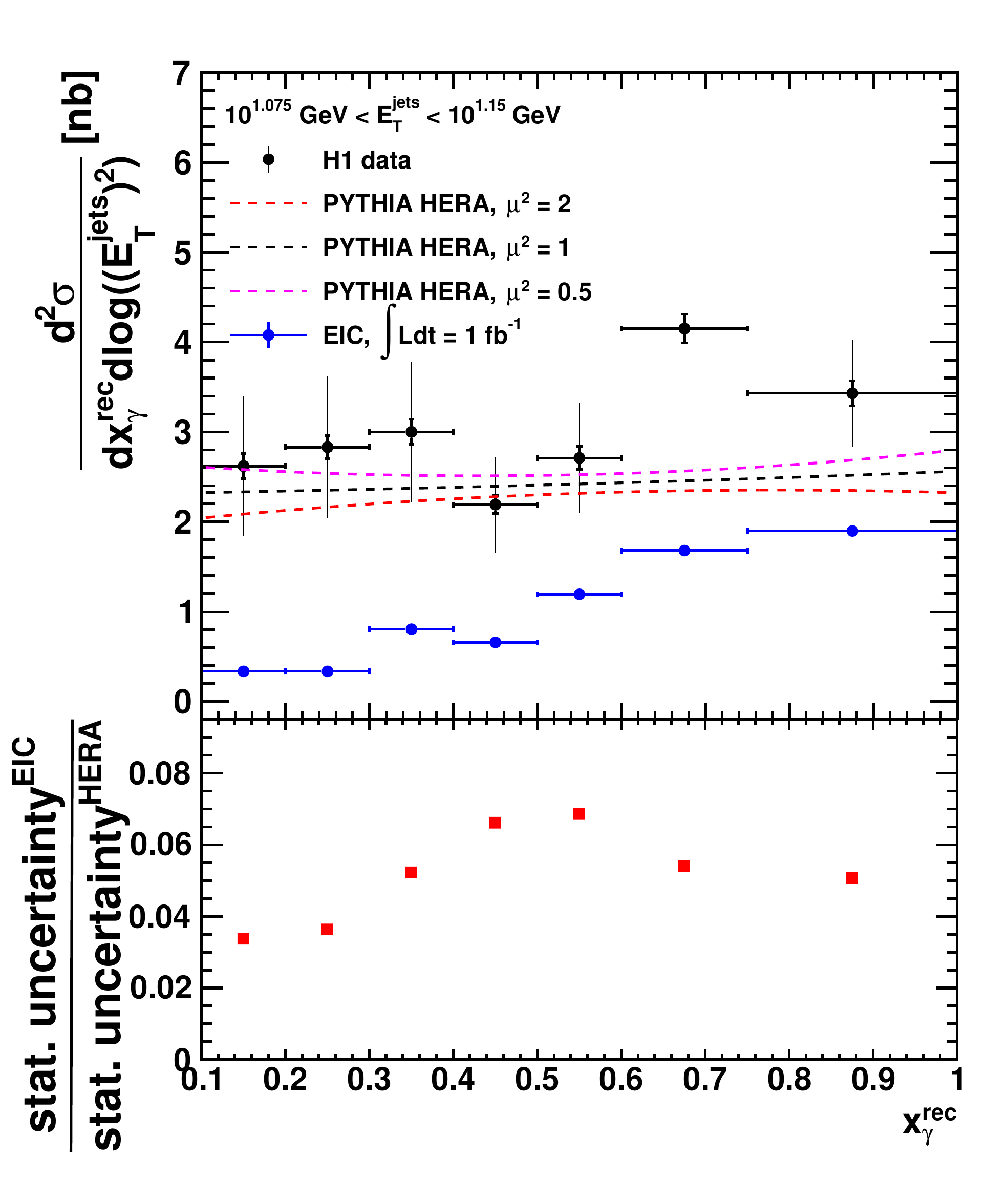}
\includegraphics[width=0.45\textwidth]{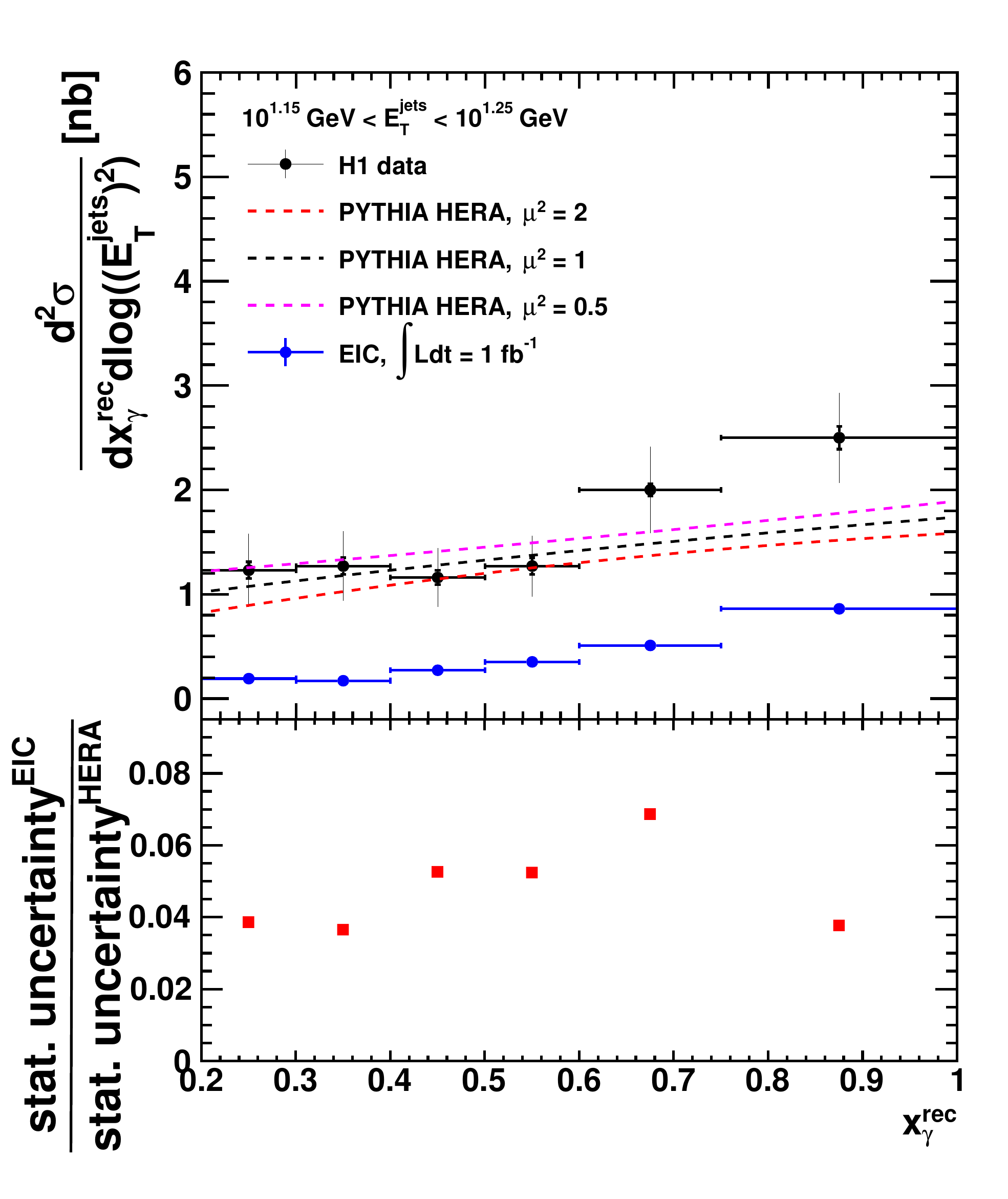}
\end{center} 
\caption{[color online] The di-jet cross section at EIC and HERA for two different 
$E_{T}^\textrm{jets}$ bins as a function of $x_{\gamma}^{rec}$. The dashed curves represent the PYTHIA-6 simulation for HERA with varying the renormalization and factorization scale $\mu^2$ 
by 0.5 (magenta) and 2 (red), 
respectively. The EIC kinematics are the same as in FIG.~\ref{fig:unxsection}. The H1 data are taken 
from~\cite{H1Collaboration6}; the inner error bars represent the statistical uncertainties, the 
outer ones are the quadratic sum of the  statistical and systematic uncertainties. 
Left: $10^{1.075} \textrm{GeV} < E_{T}^\textrm{jets} < 10^{1.15} \textrm{GeV}$. 
Right: $10^{1.15} \textrm{GeV} < E_{T}^\textrm{jets} < 10^{1.25} \textrm{GeV}$. Bottom panel: 
The ratio of the statistical uncertainties between predicted for EIC and the measured HERA data.}
\label{fig:EICadvantage}
\end{figure*}

\begin{figure*} 
\begin{center}
\includegraphics[width=0.45\textwidth]{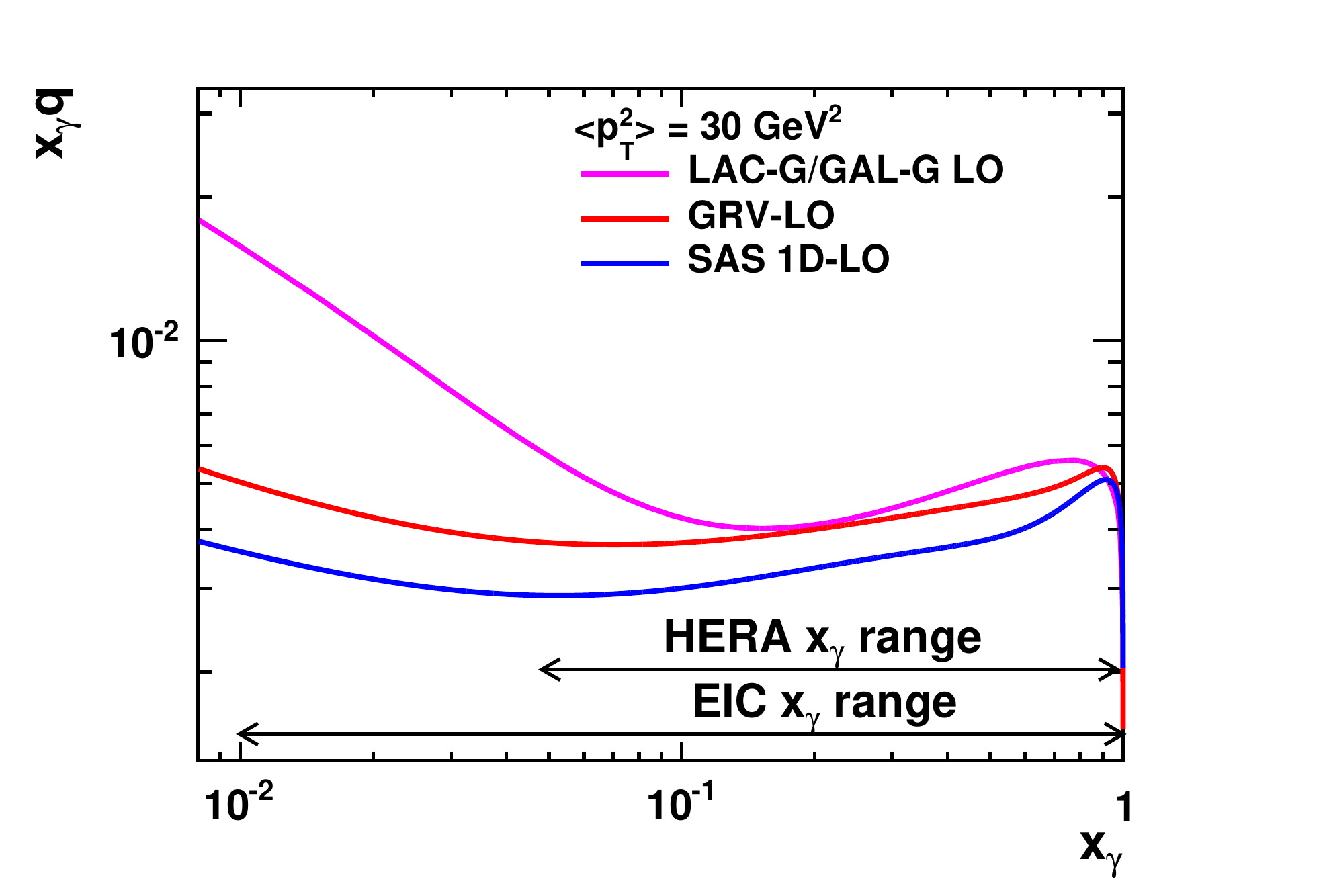}
\includegraphics[width=0.45\textwidth]{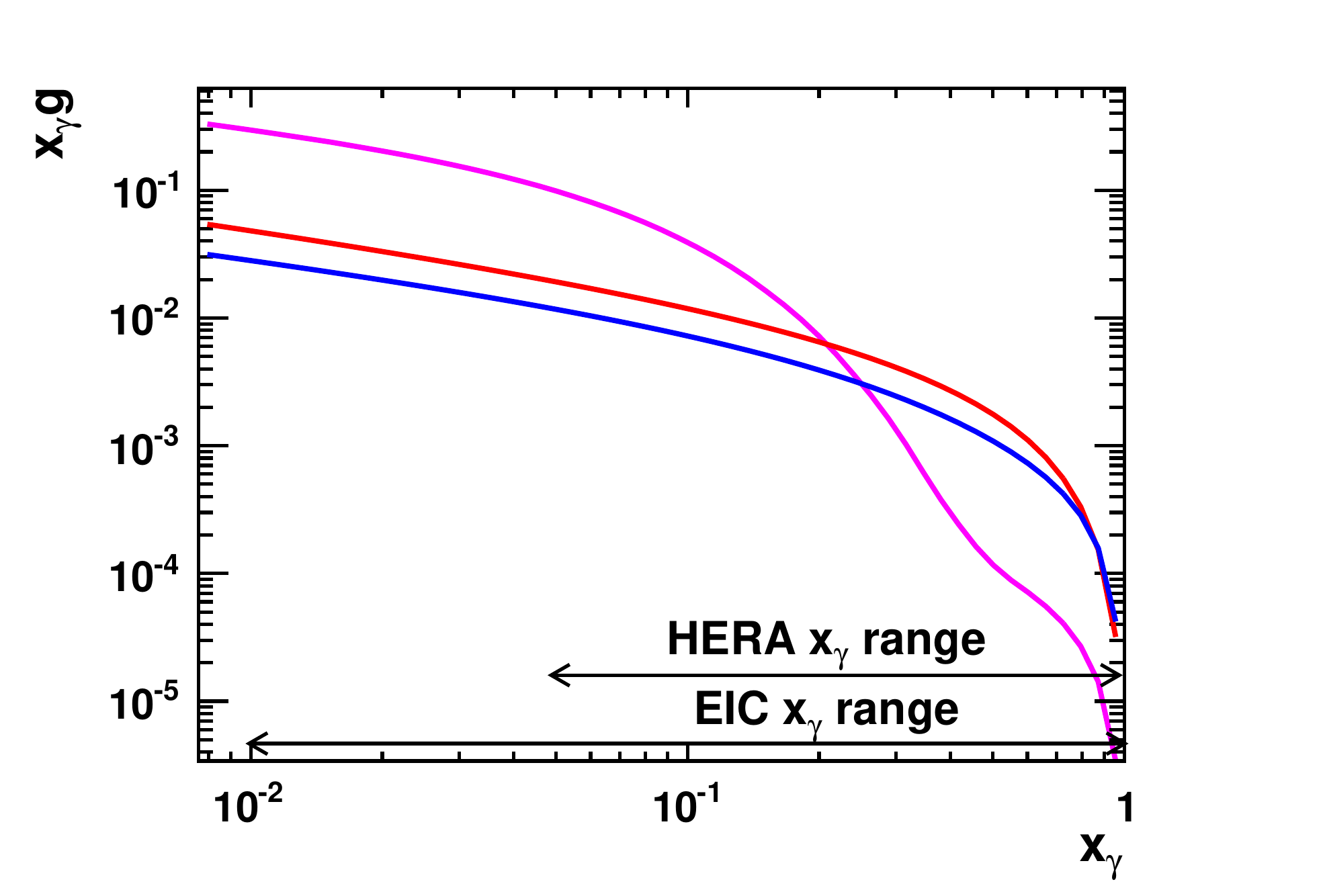}
\end{center} 
\caption{[color online] The quark (left) and gluon (right) distribution functions 
of the photon for three different sets: LAC-G (magenta), GRV-LO (red) and SAS 1D-LO (blue) at the average 
$p_{T}^{2}$ = 30 GeV$^{2}$. Indicated are the $x_{\gamma}$ ranges covered by the HERA data and as 
anticipated for the EIC data.}
\label{fig:inputPDF}
\end{figure*}

To determine the involved parton flavor of the photon, we need to use the information from the 
charged hadron with the highest $p_{T}$ (leading hadron) inside the photon side jet.
The correlation between the leading hadron type and the underlying parton flavor is shown in 
FIG.~\ref{fig:flavortagging}.
The photon side jet from a u quark in the initial state has most likely a $\pi^{+}$ as leading hadron.
Similarly, the photon side jet from an s quark is more likely to contain a leading $K^{-}$.
The sensitivity of the parton flavor to the leading hadron type can be enhanced further
with a higher leading hadron $p_{T}$ fraction ($\frac{p_{T}^{L}}{p_{T}^\textrm{{jet}}}$) cut.
This relation is particularly strong for strange quarks and kaons.

In conclusion, it is possible to tag the parton flavor of the photon by selecting the outgoing jet with the more 
negative pseudo-rapidity, and placing a cut on the leading hadron type with the requirement of the leading hadron 
carrying a high $p_{T}$ fraction.
In FIG.~\ref{fig:PKbeam parton}, we present the flavor distribution of the beam parton after selecting those jets 
with the leading hadron to be a pion or kaon.
If the leading particle of the jet is a $\pi^{+}$ ($\pi^{-}$), the most likely scenario is that this jet originated 
from a $u$ ($\bar{u}$) quark.
For $K^{+}$ ($K^{-}$), $u$ ($\bar{u}$) and $\bar{s}$ ($s$) quarks have the highest probability of being the initial 
quark. 
The photon side jet can be also initiated by gluons; a method to separate quark and gluon jets at an EIC will be 
discussed in a separate paper ~\cite{EICjet}.

Based on the results discussed in section~\ref{sec:UnpolarizedPhotonMC} and
\ref{sec:FlavorTagging}, the main advantages of an EIC to constrain the unpolarized photon PDFs can 
be summarized as:  
\begin{enumerate}
\item The existing world data both from $e^{+}e^{-}$ collisions and HERA are statistically limited. As 
described earlier, the existing photon PDFs don't provide an evaluation of their uncertainty bands, 
which makes it impossible to reach a quantitative assessment of the impact from EIC data. 
The HERA data are consistent within uncertainties with GRV-LO and SAS 1D-LO (see Figure.~7 in 
Reference~\cite{H1Collaboration5}). FIG.~\ref{fig:EICadvantage} shows a comparison of the 
statistical precision of the di-jet cross section for two bins in $E_{T}^\textrm{jets}$ 
with overlapping kinematics at HERA and EIC as a function of $x_{\gamma}^{rec}$. 
The superior statistical precision of an EIC (bottom panels) will allow a precision determination 
of the photon PDFs and their uncertainties.
The HERA data and PYTHIA-6 simulation are the same as shown in FIG.~\ref{fig:HERAcomparison}.
Also shown in FIG.~\ref{fig:EICadvantage} (top panels) is the variation of the renormalisation and 
factorization scale by 0.5 and 2; the scale dependence is small.
\item The high statistical precision will be critical to constrain the photon PDFs at 
lower $x_{\gamma}$. At EIC $x_{\gamma}>0.01$ can be reached compared with 
$x_{\gamma} > 0.05$ at HERA~\cite{H11} and $x_{\gamma} > 0.01$ for the $e^+e^-$ data~\cite{LEP3Xrange} used to 
constrain the
photon PDFs. From FIG.~\ref{fig:inputPDF}, it can be seen that the different photon PDFs diverge at 
$x_{\gamma} < 0.1$, a region where a high statistics measurement can differentiate between them.
\item The current world data do not provide any information to disentangle the different quark 
flavors. The described tagging method provides a new way to independently constrain the separate 
(anti-)quarks flavors, which is a significant step forward.
\end{enumerate} 

\subsection{The Polarized Photon Structure}
\begin{figure*} 
\begin{center}
\includegraphics[width=0.45\textwidth]{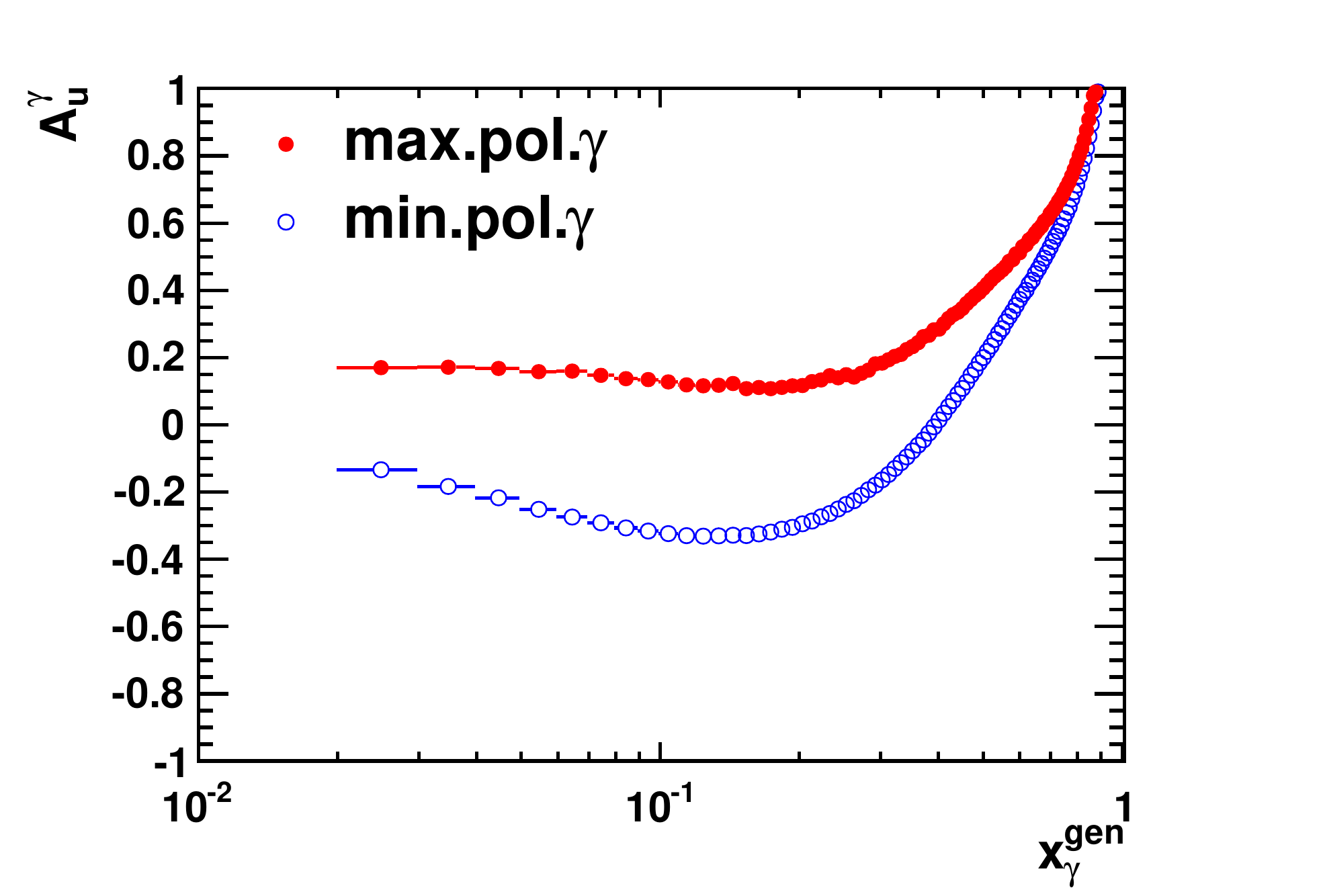}
\includegraphics[width=0.45\textwidth]{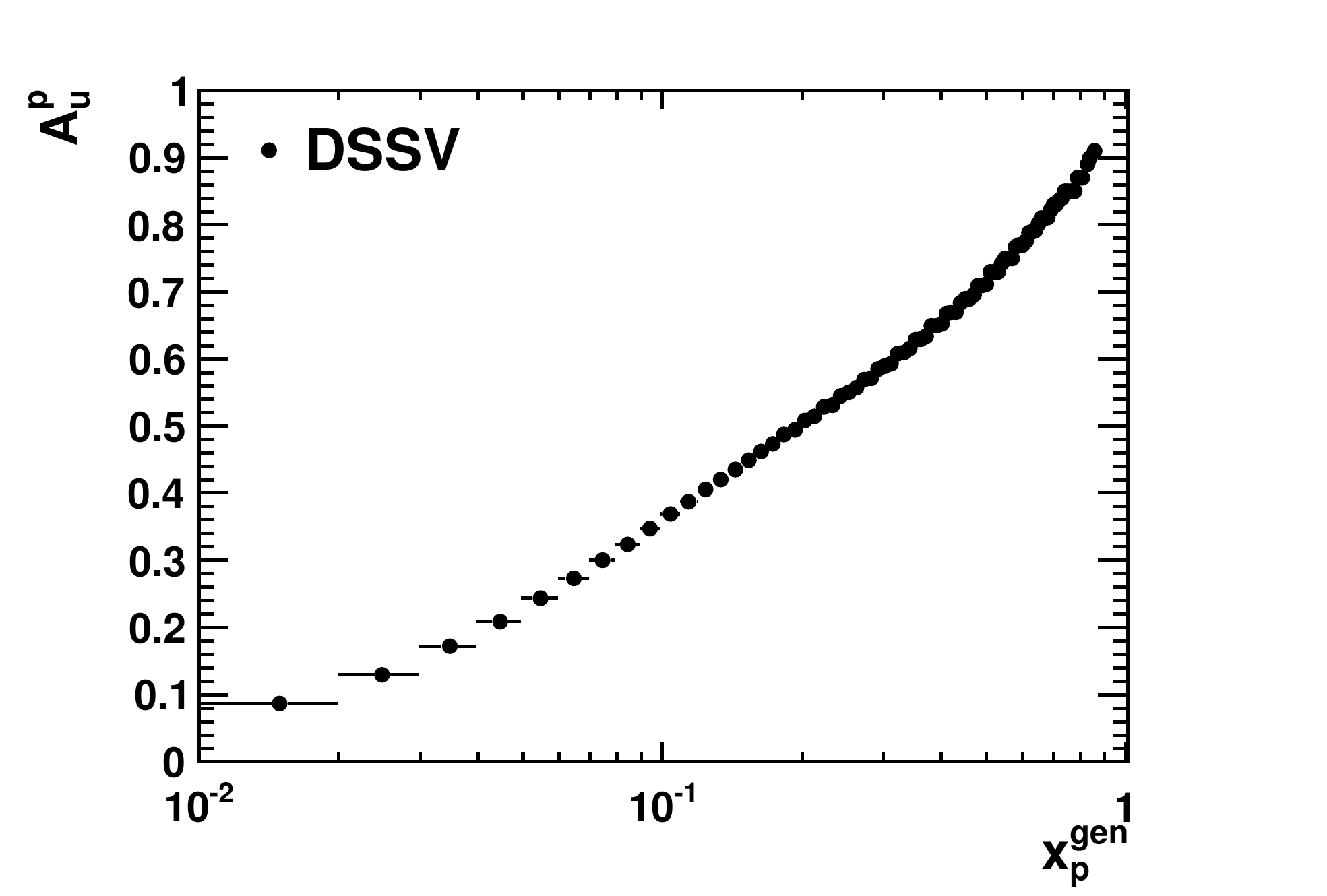}
\end{center} 
\caption{[color online] Examples of the input photon asymmetry and proton asymmetry. Left: photonic parton 
asymmetries $A_{u}^{\gamma}\equiv \Delta f^{\gamma}_{u}/f^{\gamma}_{u}$ for the ``maximal" and ``minimal" 
scenarios of the selected di-jet events at EIC.
Right: protonic parton asymmetry $A_{u}^{p}\equiv \Delta f^{p}_{u}/f^{p}_{u}$ for the DSSV polarized proton 
PDF.}
\label{fig:Aphotonproton}
\end{figure*}

The longitudinally polarized photon PDFs can be extracted measuring the polarized di-jet cross section

\begin{equation}\label{polarizedPDF}
\frac{d^{2}\Delta\sigma}{dx_{\gamma}dQ^{2}}= \Delta \gamma_{flux}\otimes \Delta f_{\gamma}(x_{\gamma},Q^{2},\mu)\\ 
\otimes \Delta f_{p}(x_{p},\mu) \otimes \sigma_{ij}
\end{equation}
with the polarized cross section defined as

\begin{equation}\label{polarizedcrosssection}
\Delta \sigma = \frac{1}{2}(\sigma(++) - \sigma(+-)) ,
\end{equation}
and $+$, $-$ denoting the helicity of the scattering particles.
$\Delta f_{\gamma}$ and $\Delta f_{p}$ represent the polarized photon PDFs and proton PDFs, respectively.
The relevant cross section asymmetry measured experimentally is $A_{LL} = \Delta \sigma / \sigma$.
Because PYTHIA does not incorporate spin dependent cross sections, this information needs to be constructed externally.
A relatively straightforward way of doing this is to calculate asymmetries on an event-by-event
basis and then apply the asymmetry as an event weight.
For this analysis, event weights were calculated depending on the kinematics and subprocesses as generated in PYTHIA,
and applied in an external analysis of the PYTHIA output.
The event information available from PYTHIA is the kinematics ($x, Q^{2}$).
The asymmetry weight for a given process can be constructed as
\begin{equation}\label{eqa:weight}
w=\hat{a}(\hat{s},\hat{t},\mu^{2},Q^{2})\times \frac{\Delta f_{a}^{\gamma}(x_{a},\mu^{2}
)}{f_{a}^{\gamma}(x_{a},\mu^{2})}\times \frac{\Delta f_{b}^{p}(x_{b},\mu^{2}
)}{f_{b}^{p}(x_{b},\mu^{2})}
\end{equation}
where $\hat a$ is the hard subprocess asymmetry.
The leading-order formulas for helicity-dependent and helicity-averaged cross sections for scattering of partons in the PGF, QCDC and DIS
subprocess are taken from~\cite{ABravar} and the lowest order equations~\cite{JBabcock} for the resolved subprocesses asymmetries
are obtained from~\cite{Hermes} . 
 
The second term of Eq.~{\ref{eqa:weight}}, $\frac{\Delta f^{\gamma}}{f^{\gamma}}$, is the photonic parton asymmetry.
$\frac{\Delta f_{a}^{\gamma}(x_{a},\mu^{2})}{f_{a}^{\gamma}(x_{a},\mu^{2})} = 1$ for $x_{a} = 1$ in direct photon processes (PGF, QCDC, DIS).
In resolved processes (hard QCD $2\rightarrow2$ processes), $f^{\gamma}$ is the unpolarized photon PDF;
we use the aforementioned SAS 1D-LO PDF.
$\Delta f^{\gamma}(x,\mu^{2})$, the parton distributions of longitudinally polarized photons, are experimentally completely unknown.
In this analysis, two very different scenarios~\cite{MGluck1,MGluck2,MGluck3,JMButterworth1} for the polarized photon PDFs were considered,
assuming ``maximal" ($\Delta f^{\gamma}(x,\mu^{2})=f^{\gamma}(x,\mu^{2}$)) or ``minimal" ($\Delta f^{\gamma}(x,\mu^{2})=0$) polarization
based on the positivity constraints
  \begin{equation}
 |\Delta f^{\gamma}(x,\mu^{2})|\leq f^{\gamma}(x,\mu^{2})
\end{equation}
at the input scale $\mu$ (also commonly referred to as $\hat{Q}^{2}$),  where $\mu$ is defined to be
  \begin{equation}
  \mu^{2}=\hat{p}_{T}^{2}+\frac{1}{2}Q^{2}
 \end{equation}
 
To take the $u$ quark as an example, the results of the two assumptions are presented in the left of FIG.~\ref{fig:Aphotonproton}
in terms of the photonic parton asymmetries $A_{f}^{\gamma}\equiv \Delta f^{\gamma}/f^{\gamma}$,
for our event selection at $Q^{2}<0.1$ $\mathrm{GeV^{2}}$ in LO.
These sets are used in the following to calculate the di-jet double spin asymmetry $A_{LL}$.
The third term of Eq.~{\ref{eqa:weight}} is the parton asymmetry in the proton, defined as $A_{f}^{p}\equiv \Delta f^{p}/f^{p}$.
In the simulation, the input for $f^{p}$ is the unpolarized proton PDF of CTEQ5m and we choose DSSV~\cite{Asymmetry,DdeFlorian}
for the polarized proton PDF; this is shown in the right of FIG.~\ref{fig:Aphotonproton}.

\begin{figure}[!h]\label{fig:polarizedxsection} 
 \centering
  \includegraphics[width=0.49\textwidth]{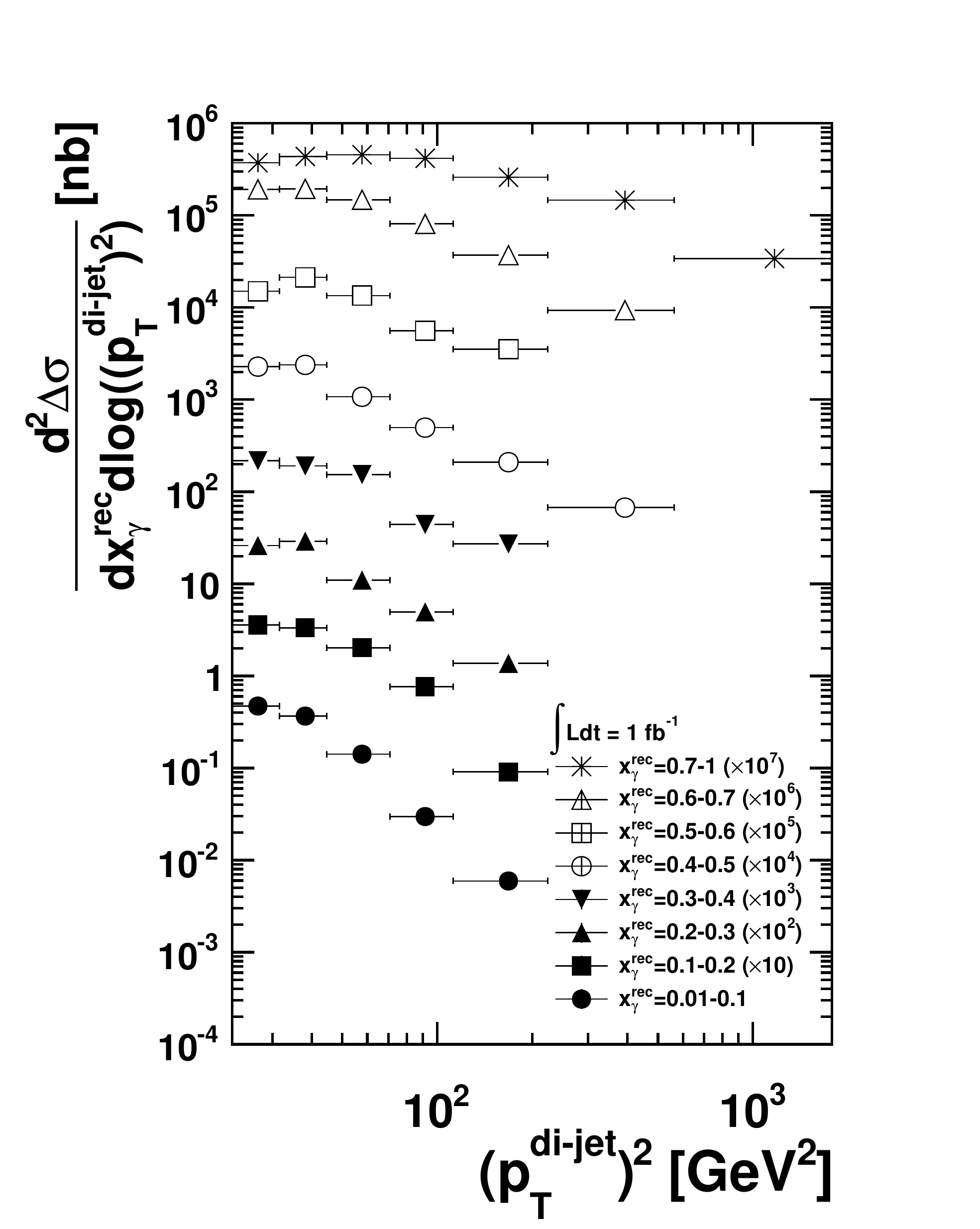}
  \caption{The measured di-jet cross section in polarized $ep$ collision as a function of the squared jet transverse momentum
  for the range of the reconstructed parton fractional momentum.
  The kinematics are the same as in FIG.~\ref{fig:unxsection}.}
  \label{fig:polarizedxsection}
\end{figure}

\begin{figure*}
\begin{center}
\includegraphics[width=0.8\textwidth]{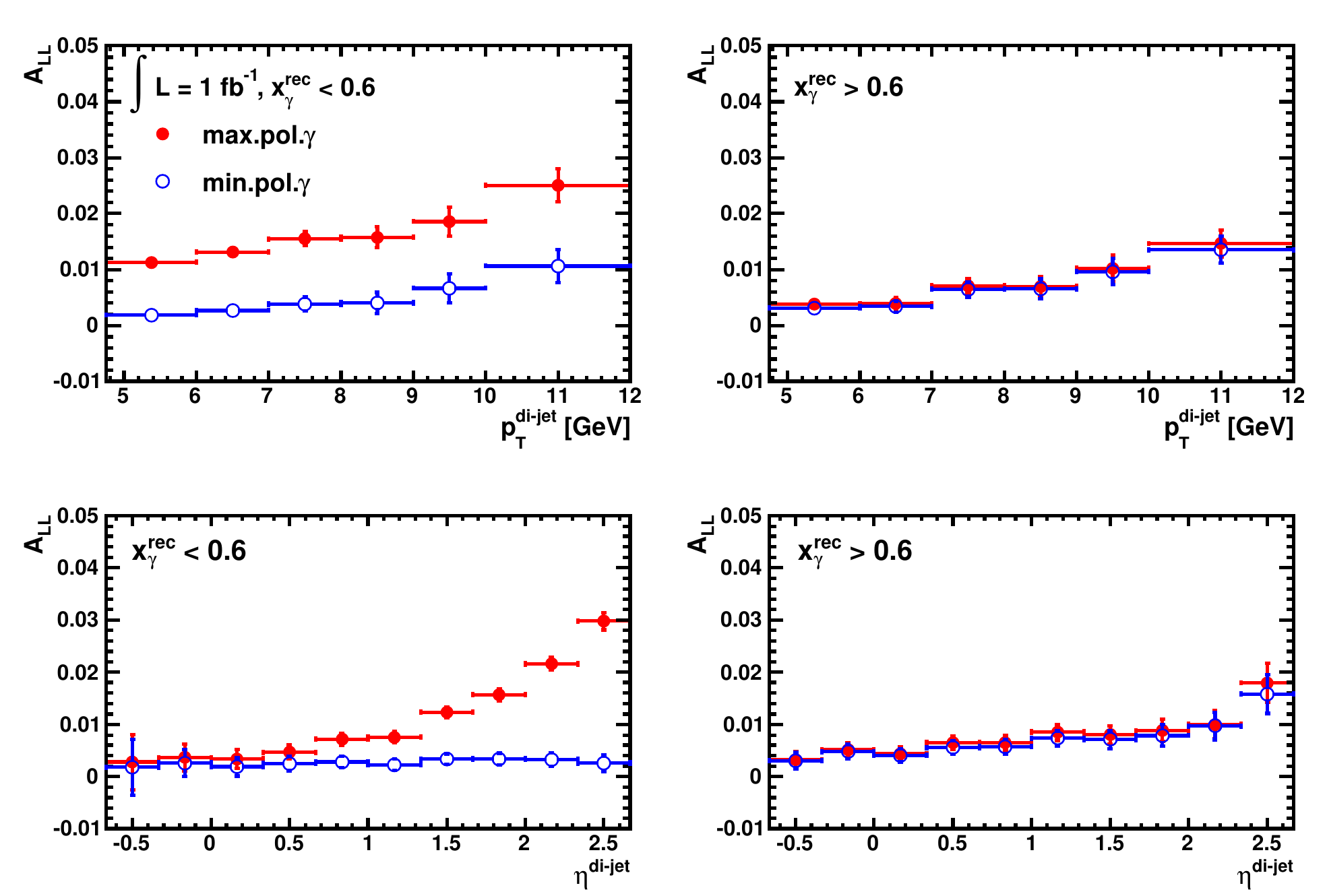} 
\end{center} 
\caption{[color online] Top: the cross section asymmetry as a function of $p_{T}^\textrm{{di-jet}}$ (top) and $\eta^\textrm{{di-jet}}$ (bottom). The two columns show the asymmetry for two $x^{rec}_{\gamma}$ regions to enhance the contribution from the resolved photon processes (left) and the direct photon processes (right). The kinematics are the same as in FIG.~\ref{fig:unxsection}.}
\label{fig:ALL} 
\end{figure*}

The final polarized di-jet cross section is measured with the same selection criteria as for the unpolarized cross section and applying the event weights as described before.
FIG.~\ref{fig:polarizedxsection} shows the polarized di-jet cross section using the ``maximal" polarization scenario for the partons in the photon.
The polarized di-jet cross section can be measured at an EIC over a wide kinematic range with high accuracy.

The statistical errors $\delta A$ are estimated from
 \begin{equation}\label{Aerror}
 \delta A = \frac{1}{\sqrt{\mathcal{L}\sigma}} ,
 \end{equation}
where the integrated luminosity $\mathcal{L} = 1 fb^{-1}$.

In FIG.~\ref{fig:ALL} $A_{LL}$ as a function of $p_{T}^\textrm{{di-jet}}$ is shown. 
In FIG.~\ref{fig:ALL} top left the asymmetry is displayed by applying a cut $x^{rec}_{\gamma}<0.6$ to select a region where the resolved processes dominate. 
The ``maximal" and ``minimal" scenario for the polarization of the partons in the photon lead to a significant difference in the predicted asymmetry.
The structure of the photon plays an important role in this region. 
Approaching the large $x^{rec}_{\gamma}$ region, the direct processes start to dominate.
In this kinematic region the asymmetry is dominated by the polarization of the partons in the proton, the photon is mainly a point-like particle, therefore the two scenarios converge to the same $A_{LL}$.
The bottom of FIG.~\ref{fig:ALL} shows $A_{LL}$ as a function of $\eta^\textrm{{di-jet}}$.
The overall behavior of the asymmetry as a function of $\eta^\textrm{{di-jet}}$ follows the one as a function of $p_{T}^\textrm{{di-jet}}$, showing a significant difference in the $x^{rec}_{\gamma}$ region where the resolved processes dominate.
In conclusion, the cross section asymmetry is sensitive to the polarization of the partons in the photon in the resolved photon processes, and the polarized photon PDFs can be well constrained by measuring $A_{LL}$ at an EIC.

\section{Summary}\label{sec:summary}
The hadronic structure of the photon can be accessed at low $Q^{2}$ in deep inelastic scattering through tagging resolved photon processes.
We have shown in a detailed analysis the capability of a future EIC to perform di-jet measurements to extract (un)polarized photon PDFs:
di-jets produced in direct and resolved process can be well separated by reconstructing $x_{\gamma}$,
which has a strong correlation with the true $x_{\gamma}$,
and one can effectively extract the underlying photon PDFs by measuring di-jet cross sections in photoproduction events.
Jets from the photon side can be identified by selecting the more negative pseudo-rapidity jet in each event.
Moreover, it is possible to probe the content of the photon by tagging leading hadrons inside the jets from the photon side;
the flavor of the originating quark is highly correlated with the identified hadron.
With polarized beams, the polarized photon PDFs, which are totally unknown so far, can be extracted at an EIC. 

\begin{acknowledgments}
We are very grateful to the EIC group at BNL whose ongoing efforts made this analysis possible.
We acknowledge the BNL EIC task force for effective discussions. We appreciate the guiding suggestions from M. Stratmann. And we acknowledge the advice from W.B. Schmidke. 
E.C.A. and J.H.L. acknowledge the support by the U.S. Department of Energy under
contract number No. DE-SC0012704.
This work is also supported by the program of Introducing Talents of Discipline to Universities (B08033),
the NSFC (11475068) and the National Key Research and Development Program of China (2016YFE0100900).
\end{acknowledgments}


\newpage
\begin{thebibliography}{99}
\bibitem{Planck} 
 Annalen der Physik. 4 (3): 553 (1901)
 [doi:10.1002/andp.19013090310]

\bibitem{EinsteinA} 
 A.~Einstein,
   Annalen der Physik. 17 (6): 132-148 (1905)
   
 \bibitem{VVKobychev} 
 V.~V.~Kobychev and S.~B.~Popov,
  Astron.\ Lett.\ {\bf 31}, 147-151 (2005)
  
  \bibitem{WalshTF} 
 T.F.~Walsh and P.~Zerwas,
  Phys.\ Lett.\ B {\bf 44} 195-198 (1973)
  
  \bibitem{WSlominski} 
W.~Slominski and J.~Szwed,
  Acta Phys.\ Polon. {\bf 27}, 1887-1914 (1996)
[arXiv:hep-ph/9606395]
  
  \bibitem{WittenE} 
 E.~Witten,
  Nucl.\ Phys.\ B {\bf 120}, 189-202 (1977)

\bibitem{F2}
PLUTO Collab., Ch.~Berger et al., Phys.\ Lett.\ B {\bf142} (1984) 111, Nucl.\ Phys.\ B {\bf 281} (1987) 365, Phys.\ Lett.\ B {\bf107} (1981) 168, Phys.\ Lett.\ B {\bf 142} (1984) 119;
JADE Collab., W.~Bartel et al., Z.\ Phys.\ C {\bf 24} (1984) 231;
 TASSO Collab., M.~Althoff et al., Z.\ Phys.\ C {\bf 24} (1984) 231;
 TPC/2$\gamma$ Collab., H.~Aihara et al., Phys.\ Rev.\ Lett. 58 (1987) 97, Z.\ Phys.\ C {\bf 34} (1987) 1;
 AMY Collab., T.~Sasaki et al., Phys. Lett.\ B {\bf 252} (1990) 491, Phys.\ Lett.\ B {\bf 325} (1994) 248;
 OPAL Collab., R.~Akers et al., Z.\ Phys.\ C {\bf 61} (1994) 199;
 OPAL collab., Phys.\ Lett.\ B {\bf 539} 13-24 (2002).
 OPAL Collab., Phys.\ Lett.\ B {\bf 412} 225-234 (1997).
 TOPAZ Collab., H.~Hayashii et al., Phys.\ Lett.\ B {\bf 314} (1993) 149;
 ALEPH Collab., D.~Buskulic et al., Phys.\ Lett.\ B {\bf 313} (1993) 509;
 DELPHI Collab., P.~Abreu et al., Phys.\ Lett.\ B {\bf 342} (1995) 402
  
 \bibitem{ArminBöhrer} 
 Armin~B\"ohrer,
   [arXiv:hep-ex/0305029]
  
  \bibitem{ADeRoeck}
   A.~De~Roeck,
    Eur.\ Phys.\ J.\ C {\bf 33} s394-s397 (2004)
    
   \bibitem{JMButterworth}
  J.~M.~Butterworth,
  DESY 95-043
    [arXiv:hep-ex/9503011]
   
  \bibitem{PJBussey}
P.~J.~Bussey,
Nucl.\ Phys.\ Proc.\ Suppl.{\bf 126} 17-21(2004)
    
 \bibitem{H1Collaboration1}
H1 Collaboration,
Phys.\ Lett.\ B {\bf 314} 436 (1993),
Phys.\ Lett.\ B {\bf 368} 412-422 (1995), Nucl.\ Phys.\ B {\bf 445} (1995) 195,
Eur.\ Phys.\ J.\ C {\bf 13} 397-414 (2000), Nucl.\ Phys.\ B {\bf 82} (2000) 118
Eur.\ Phys.\ J.\ C {\bf 25} (2002) 13-23,

    ZEUS Collaboration,
Phys.\ Lett.\ B {\bf 348} 665-680 (1995),
Phys.\ Lett.\ B {\bf 384} 401 (1996),
Eur.\ Phys. \ J.\ C {\bf 4} 591 (1998)
    
  \bibitem{H1Collaboration5}
H1 Collaboration,
Eur.\ Phys. \ J.\ C {\bf 10} 363-372 (1999) 
    
    \bibitem{SFrixione}
S.~Frixione, P.~Nason and G.~Ridolfi,
Nucl.\ Phys.\ B {\bf 454} 3-24 (1995) 
 
   \bibitem{AAccardi}
A.~Accardi $et~~al.,$
 Eur.\ Phys.\ J.\ A {\bf 52} (2016) 268 
 
 \bibitem{ECAschenauer}
E.C.~Aschenauer $et~~al.,$
 [arxiv: 1409.1633] 
 
  \bibitem{Sjostrand}
  T.~Sjostrand, S.~Mrenna and P.~Z.~Skands,
  JHEP {\bf 0605}, 026 (2006)
  
   \bibitem{Whalley} 
  M.~R.~Whalley, D.~Bourilkov and R.~C.~Group,
  [arXiv:hep-ph/0508110]
 
  \bibitem{CTEQ5}
  H.~L.~Lai, J.~Huston, S.~Kuhlmann, J.~Morfin, F.~Olness, J.~F.~Owens, J.~Pumplin and W.~K.~Tung, 
  Eur.\ Phys.\ J.\ C {\bf 12} (2000) 375-392
  
   \bibitem{IAbt2} 
  I.~Abt, A.M.~Cooper-Sarkar, B.~Foster, V.~Myronenko, K.~Wichmann and  M. ~Wing,
Phys.\ Rev.\ D {\bf 96} 014001 (2017)
  
  \bibitem{SAS}
  Gerhard A.~Schuler and Torbjoern~Sjoestrand,
  Zeit.\ Phys.\ C {\bf 68} (1995) 607
  
 
 \bibitem{DG}
 M.~Dress and K.~Grassie,
 Z.\ Phys.\ C {\bf 28} (1985) 451
 
 \bibitem{LAC}
 H.~Abramowicz, K.~Charchuta and A.~Levy,
Phys.\ Lett.\ B269 (1991) 458-464

\bibitem{GS}
L.~E~Gordon and J.~K.~Storrow,
Z.\ Phys.\ C {\bf 56} (1992) 307

\bibitem{GS96}
L.~E~Gordon and J.~K.~Storrow,
Nucl.\ Phys.\ B {\bf 489} (1997) 405

\bibitem{GRV}
M.~Gluck, E.~Reya and A.~Vogt,
Phys.\ Rev.\ D {\bf 45} (1992) 3986, Phys.\ Rev.\ D {\bf 46} (1992) 1973;
M.~Gluck, E.~Reya and M.~Stratmann,
Phys.\ Rev.\ D {\bf 51} (1995) 3220

\bibitem{ACFGP}
P.~Aurenche, J.-P.~Guillet, M.~Fontannaz, 
Z.\ Phys.\ C {\bf 64} (1994) 621;
P.~Aurenche, P.~Chiappetta, M.~Fontannaz, J.P.~Guillet and E.~Pilon,
Z.\ Phys.\ C {\bf 56} (1992) 589

\bibitem{SAS1}
Gerhard~A.~Schuler and Torbjoern~Sjoestrand,
Phys.\ Lett.\ B {\bf 376} (1996) 193

  \bibitem{DO}
   D.~W.~Duke and J.~F.~Owens,
   Phys.\ Rev.\ D {\bf 26} (1982) 1600-1609

\bibitem{H1}
H1 Collaboration,
Phys.\ Lett.\ B {\bf 25} (1997) 418

\bibitem{EICjet}
X. Chu, B. Page and E.C. Aschenauer in preparation

\bibitem{H11}
H1 Collaboration,
Phys.\ Lett.\ B {\bf 483} (2000) 36–48

\bibitem{ZEUS}
ZEUS Collaboration, 
Eur.\ Phys.\ J.\ C {\bf 23} (2002) 615, Phys. \ Lett.\ B {\bf 354} (1995) 163.

\bibitem{MCacciari} 
 M.~Cacciari and G.P.~Salam,
 Phys.\ Lett.\ B {\bf 641} 57 (2006)

\bibitem{H1Collaboration6}
H1 Collaboration,
Eur.\ Phys.\ J.\ C {\bf 1} 97-107 (1998) 
    
\bibitem{MCacciari1}
M.~Cacciari, G.P.~Salam and G.~Soyez,
 JHEP {\bf 0804} 063 (2008)

\bibitem{Jacquet}
F.~Jacquet and A.~Blondel,
Report From The Study Group On Detectors For Charged Current Events - in: U. Amaldi et al DESY, Hamburg, 1979, pp. 391;  For a recent overview of different methods to determine kinematic variables in DIS see: J. Blumlein, Prog. Part. Nucl.\ Phys. {\bf 69} (2013) 28

\bibitem{elke}
Elke C.~Aschenauer, Thomas Burton, Till Martini, Hubert Spiesberger, and Marco Stratmann,
Phys.\ Rev.\ D {\bf 88} (2013) 114025

\bibitem{TMDs}
Zhong-Bo Kang, Alexei Prokudin, Felix Ringer and Feng Yuan, 
[arXiv:1707.00913];
Zhong-Bo Kang, Xiaohui Liu, Felix Ringer and Hongxi Xing,
[arXiv:1705.08443]

\bibitem{FFs}
Tom Kaufmann, Asmita Mukherjee, and Werner Vogelsang,
Phys.\ Rev.\ D {\bf 92} (2015) 054015,
Phys.\ Rev.\ D {\bf 93} (2016) 114021

\bibitem{ATLASFFs}
The ATLAS Collaboration,
[arXiv:1706.02859]

\bibitem{STAR}
J.K.~Adkins,
Int.\ J.\ Mod.\ Phys.\ Conf.\ Ser. {\bf 40} (2016) 1660040,
PoS DIS2015 (2015) 193

\bibitem{LEP3Xrange}
L3 Collaboration,
Phys.\ Lett.\ B {\bf 447} (1999) 147-156
    
\bibitem{ABravar}
A.~Bravar, K.~Kurek and R.~Windmolders,
Comput.\ Phys.\ Commun. {\bf 105} (1997) 42-61
    
\bibitem{JBabcock}
J.~Babcock, E.~Monsay and D.~W.~Sivers,
Phys.\ Rev.\ D {\bf 19} (1979) 1483

\bibitem{Hermes}
The Hermes Collaboration, Airapetian, A., Akopov, N. et al,
J.\ High\ Energ.\ Phys. (2010) 2010: 130 
    
\bibitem{MGluck1}
M.~Gluck and W.~Vogelsang
Z.\ Phys.\ C {\bf55} 353-356 (1992)

\bibitem{MGluck2}
M.~Gluck and W.~Vogelsang
Z.\ Phys.\ C {\bf57} 309-310 (1993) 

\bibitem{MGluck3}
M.~Gluck, M.~Stratmann and W.~Vogelsang
Phys.\ Lett.\ B {\bf 337} 373-375 (1994) 

\bibitem{JMButterworth1}
J.M.~Butterworth, N.~Goodman, M.~Stratmann and W.~Vogelsang
CERN-TH/97-310, DTP/97/98, UCL/HEP 97-08 (1997) 
[arXiv:hep-ph/9711250]

\bibitem{Asymmetry}
Asymmetry Analysis Collaboration
Phys.\ Rev.\ D {\bf 69} 054021 (2004)  

\bibitem{DdeFlorian}
D.~de~Florian, R.~Sassot, M.~Stratmann and W.~Vogelsang
Phys.\ Rev.\ D {\bf 80} 034030 (2009) 

\end{thebibliography}

\end{document}